\documentclass[twocolumn]{aastex62}
\usepackage{longtable}
\usepackage{amsmath}
\usepackage{subfigure}
\usepackage{appendix}

\usepackage{graphicx}
\usepackage{dcolumn}
\usepackage{bm}

\renewcommand*{\thefootnote}{\fnsymbol{footnote}}

\graphicspath{{./}{figures/}}

\begin{document}

\reportnum{\footnotesize CERN-TH-2022-216}

\title{Search for Ultraheavy Dark Matter from Observations of Dwarf Spheroidal Galaxies with VERITAS}

\date{\today}
\correspondingauthor{Donggeun Tak (\href{mailto:donggeun.tak@gmail.com}{donggeun.tak@gmail.com}), Elisa Pueschel (\href{mailto:elisa.pueschel@desy.de}{elisa.pueschel@desy.de})}

\author{A.~Acharyya}\affiliation{Department of Physics and Astronomy, University of Alabama, Tuscaloosa, AL 35487, USA}
\author{A.~Archer}\affiliation{Department of Physics and Astronomy, DePauw University, Greencastle, IN 46135-0037, USA}
\author{P.~Bangale}\affiliation{Department of Physics and Astronomy and the Bartol Research Institute, University of Delaware, Newark, DE 19716, USA}
\author{J.~T.~Bartkoske}\affiliation{Department of Physics and Astronomy, University of Utah, Salt Lake City, UT 84112, USA}
\author{P.~Batista}\affiliation{Deutsches Elektronen-Synchrotron DESY, Platanenallee 6, 15738 Zeuthen, Germany}
\author{M.~Baumgart}
\affiliation{Department of Physics, Arizona State University, Tempe, AZ 85287, USA}
\author{W.~Benbow}\affiliation{Center for Astrophysics $|$ Harvard \& Smithsonian, Cambridge, MA 02138, USA}
\author{J.~H.~Buckley}\affiliation{Department of Physics, Washington University, St. Louis, MO 63130, USA}
\author{A.~Falcone}\affiliation{Department of Astronomy and Astrophysics, 525 Davey Lab, Pennsylvania State University, University Park, PA 16802, USA}
\author{Q.~Feng}\affiliation{Center for Astrophysics $|$ Harvard \& Smithsonian, Cambridge, MA 02138, USA}
\author{J.~P.~Finley}\affiliation{Department of Physics and Astronomy, Purdue University, West Lafayette, IN 47907, USA}
\author{G.~M.~Foote}\affiliation{Department of Physics and Astronomy and the Bartol Research Institute, University of Delaware, Newark, DE 19716, USA}
\author{L.~Fortson}\affiliation{School of Physics and Astronomy, University of Minnesota, Minneapolis, MN 55455, USA}
\author{A.~Furniss}\affiliation{Department of Physics, California State University - East Bay, Hayward, CA 94542, USA}
\author{G.~Gallagher}\affiliation{Department of Physics and Astronomy, Ball State University, Muncie, IN 47306, USA}
\author{W.~F.~Hanlon}\affiliation{Center for Astrophysics $|$ Harvard \& Smithsonian, Cambridge, MA 02138, USA}
\author{O.~Hervet}\affiliation{Santa Cruz Institute for Particle Physics and Department of Physics, University of California, Santa Cruz, CA 95064, USA}
\author{J.~Hoang}\affiliation{Santa Cruz Institute for Particle Physics and Department of Physics, University of California, Santa Cruz, CA 95064, USA}
\author{J.~Holder}\affiliation{Department of Physics and Astronomy and the Bartol Research Institute, University of Delaware, Newark, DE 19716, USA}
\author{T.~B.~Humensky}\affiliation{Department of Physics, University of Maryland, College Park, MD, USA and }
\author{W.~Jin}\affiliation{Department of Physics and Astronomy, University of Alabama, Tuscaloosa, AL 35487, USA}
\author{P.~Kaaret}\affiliation{Department of Physics and Astronomy, University of Iowa, Van Allen Hall, Iowa City, IA 52242, USA}
\author{M.~Kertzman}\affiliation{Department of Physics and Astronomy, DePauw University, Greencastle, IN 46135-0037, USA}
\author{M.~Kherlakian}\affiliation{Deutsches Elektronen-Synchrotron DESY, Platanenallee 6, 15738 Zeuthen, Germany}
\author{D.~Kieda}\affiliation{Department of Physics and Astronomy, University of Utah, Salt Lake City, UT 84112, USA}
\author{T.~K.~Kleiner}\affiliation{Deutsches Elektronen-Synchrotron DESY, Platanenallee 6, 15738 Zeuthen, Germany}
\author{N.~Korzoun}\affiliation{Department of Physics and Astronomy and the Bartol Research Institute, University of Delaware, Newark, DE 19716, USA}
\author{F.~Krennrich}\affiliation{Department of Physics and Astronomy, Iowa State University, Ames, IA 50011, USA}
\author{M.~J.~Lang}\affiliation{School of Natural Sciences, University of Galway, University Road, Galway, H91 TK33, Ireland}
\author{M.~Lundy}\affiliation{Physics Department, McGill University, Montreal, QC H3A 2T8, Canada}
\author{G.~Maier}\affiliation{Deutsches Elektronen-Synchrotron DESY, Platanenallee 6, 15738 Zeuthen, Germany}
\author{C.~E~McGrath}\affiliation{School of Physics, University College Dublin, Belfield, Dublin 4, Ireland}
\author{P.~Moriarty}\affiliation{School of Natural Sciences, University of Galway, University Road, Galway, H91 TK33, Ireland}
\author{S.~O'Brien}\affiliation{Physics Department, McGill University, Montreal, QC H3A 2T8, Canada}
\author{R.~A.~Ong}\affiliation{Department of Physics and Astronomy, University of California, Los Angeles, CA 90095, USA}
\author{K.~Pfrang}\affiliation{Deutsches Elektronen-Synchrotron DESY, Platanenallee 6, 15738 Zeuthen, Germany}
\author{M.~Pohl}\affiliation{Institute of Physics and Astronomy, University of Potsdam, 14476 Potsdam-Golm, Germany}\affiliation{Deutsches Elektronen-Synchrotron DESY, Platanenallee 6, 15738 Zeuthen, Germany}
\author{E.~Pueschel}\affiliation{Deutsches Elektronen-Synchrotron DESY, Platanenallee 6, 15738 Zeuthen, Germany}
\author{J.~Quinn}\affiliation{School of Physics, University College Dublin, Belfield, Dublin 4, Ireland}
\author{K.~Ragan}\affiliation{Physics Department, McGill University, Montreal, QC H3A 2T8, Canada}
\author{P.~T.~Reynolds}\affiliation{Department of Physical Sciences, Munster Technological University, Bishopstown, Cork, T12 P928, Ireland}
\author{E.~Roache}\affiliation{Center for Astrophysics $|$ Harvard \& Smithsonian, Cambridge, MA 02138, USA}
\author{N.~L.~Rodd}
\affiliation{Theoretical Physics Department, CERN, 1 Esplanade des Particules, CH-1211 Geneva 23, Switzerland}
\author{J.~L.~Ryan}\affiliation{Department of Physics and Astronomy, University of California, Los Angeles, CA 90095, USA}
\author{I.~Sadeh}\affiliation{Deutsches Elektronen-Synchrotron DESY, Platanenallee 6, 15738 Zeuthen, Germany}
\author{L.~Saha}\affiliation{Center for Astrophysics $|$ Harvard \& Smithsonian, Cambridge, MA 02138, USA}
\author{M.~Santander}\affiliation{Department of Physics and Astronomy, University of Alabama, Tuscaloosa, AL 35487, USA}
\author{G.~H.~Sembroski}\affiliation{Department of Physics and Astronomy, Purdue University, West Lafayette, IN 47907, USA}
\author{R.~Shang}\affiliation{Department of Physics and Astronomy, University of California, Los Angeles, CA 90095, USA}
\author{M.~Splettstoesser}\affiliation{Santa Cruz Institute for Particle Physics and Department of Physics, University of California, Santa Cruz, CA 95064, USA}
\author{D.~Tak}\affiliation{Deutsches Elektronen-Synchrotron DESY, Platanenallee 6, 15738 Zeuthen, Germany}\affiliation{SNU Astronomy Research Center, Seoul National University, 1 Gwanak-rho, Gwanak-gu, Seoul, Korea}
\author{J.~V.~Tucci}\affiliation{Department of Physics, Indiana University-Purdue University Indianapolis, Indianapolis, IN 46202, USA}
\author{V.~V.~Vassiliev}\affiliation{Department of Physics and Astronomy, University of California, Los Angeles, CA 90095, USA}
\author{D.~A.~Williams}\affiliation{Santa Cruz Institute for Particle Physics and Department of Physics, University of California, Santa Cruz, CA 95064, USA}

\begin{abstract}
Dark matter is a key piece of the current cosmological scenario, with weakly interacting massive particles (WIMPs) a leading dark matter candidate. WIMPs have not been detected in their conventional parameter space (100 GeV $\lesssim M_{\chi} \lesssim$ 100 TeV), a mass range accessible with current Imaging Atmospheric Cherenkov Telescopes. As ultraheavy dark matter (UHDM; $M_{\chi} \gtrsim$ 100 TeV) has been suggested as an under-explored alternative to the WIMP paradigm, we search for an indirect dark matter annihilation signal in a higher mass range (up to 30 PeV) with the VERITAS gamma-ray observatory. With 216 hours of observations of four dwarf spheroidal galaxies, we perform an unbinned likelihood analysis. We find no evidence of a $\gamma$-ray signal from UHDM annihilation above the background fluctuation for any individual dwarf galaxy nor for a joint-fit analysis, and consequently constrain the velocity-weighted annihilation cross section of UHDM for dark matter particle masses between 1 TeV and 30 PeV. We additionally set constraints on the allowed radius of a composite UHDM particle.
\end{abstract}

\section{Introduction} \label{sec:intro}
Ultra-heavy dark matter (UHDM) presents an alternative mass range for dark matter, and is partly motivated by the absence of a dark matter signature in the well-explored mass ranges suggested by the simplest dark matter models. Most dark matter searches have focused on the mass range of sub-eV (e.g., axion-like particles) or $\sim$GeV-TeV scales (e.g., weakly interacting massive particles). However, the dark matter particle mass is simply not limited to such ranges; indeed, there are many viable UHDM candidates (for a brief introduction, see \citealp{snowmass2022}). If dark matter emerged as a thermal relic from the early Universe, then as unitarity places an upper bound on its annihilation into Standard Model (SM) particles, this naively prohibits masses above $\mathcal{O}$(100) TeV if dark matter is a point-like particle~\citep{Griest1990}. Roughly, unitarity sets a maximal value for the dark matter annihilation cross section, and for dark matter heavier than $\sim 100~{\rm TeV}$, even the largest allowed cross section is insufficient to reduce the equilibrium abundance of dark matter to the observed value. However, if dark matter is made of composite states with geometrical cross sections (i.e. the cross section scales as $\pi R^2$, where $R$ is the intrinsic size of the dark matter particle), the limit is easily evaded \citep[e.g.][]{Harigaya2016, Geller2018}. One can also consider scenarios where the dark matter is not a simple thermal relic, with or without compositeness \citep[e.g.][]{Berlin2016, Contino2019}. As discussed in \cite{Tak2022}, the annihilation of UHDM particles can produce a $\gamma$-ray signal in the form of mono-energetic $\gamma$-ray lines in addition to a continuum contribution of photons with energy equal to and below the dark matter particle mass ($E_{\gamma} \lesssim M_{\chi}$), with the exact spectrum determined by the particle physics underlying the annihilation. Given this, the authors demonstrated that current very-high-energy (VHE; $\geq$ 100 GeV) $\gamma$-ray observatories are sensitive to an annihilation signal from UHDM, for masses up to at least a few tens of PeV. 

Among the best targets for indirect dark matter searches are dwarf spheroidal galaxies (dSphs) of the Local Group (located $\sim$20 to 200 kpc from Earth). Since they are dark matter-rich regions without known nearby VHE sources\footnote{A notable expection is the Sagittarius dSph: a recent study on the Fermi bubbles by \cite{Crocker2022} found a possible $\gamma$-ray signal from this dSph, attributable to millisecond pulsars.}, they have been widely studied with current VHE observatories \citep[e.g.,][]{dm_magic, dm_hawc, dm_hess, dm_hawc2, dm_magic2}. For instance, the Very Energetic Radiation Imaging Telescope Array System (VERITAS) observed five dSphs and provided upper limits on the dark matter velocity-weighted annihilation cross section in the mass range from 100 GeV to 100 TeV \citep{dm_veritas}.

In this work, we revisit the VERITAS observations of four of the five dSphs (Segue 1, Ursa Minor, Bo\"otes, and Draco; in total 216 hours of observations) and search for the indirect UHDM signal up to a mass of 30 PeV. The observation times for the targets are listed in Table~\ref{tab:nfw}. Note that we consider dSphs for which we have an estimate of the dark matter density profile; the Willman observation included in \citealp{dm_veritas} is excluded. We derive upper limits on the UHDM velocity-weighted annihilation cross section from a joint-fit maximum likelihood estimation (MLE) analysis. We further interpret the derived limits in terms of the allowed radius of a composite UHDM particle.

\begin{table*}[t!]
	\centering 
	\begin{tabular}{c | c c c c c | c c c c c | c c }
    \hline\hline
     & N$_{\rm on}$ & N$_{\rm off}$ & $\alpha$ & t$_{\rm obs}$ & $\sigma$  & $\rho_s$ & $r_s$ & $\alpha$ & $\beta$ & $\gamma$ & $\theta_{\rm max}$ & $J(\theta_{\rm max})$ \\ 
    & & & & [hrs] & & [$M_\odot/{\rm pc}^3$] & [pc] & & & & [deg] &  [${\rm GeV}^2/{\rm cm}^{5} \cdot {\rm sr}$] \\ \hline
    Segue 1 & 15895 & 120826 & 0.131 & 92.0 & 0.7 & $1.78$& $3.1\times10^2$ & 0.54 & 4.36 & 0.64 & 0.35 & $2.5\times10^{19}$\\
    Ursa Minor & 4181 & 35790 & 0.119 & 60.4 & -0.1 & $5.6\times10^{-1}$& $3.6\times10^2$ & 2.37 & 8.77 & $1.2\times10^{-2}$ & 1.19 & $7.1\times10^{18}$\\
    Bo\"otes & 1206 & 10836 & 0.116 & 14.0 & -1.0 & $6.7\times10^{-4}$& $1.2\times10^4$ & 2.81 & 4.87 & 1.08 & 0.47 & $1.7\times10^{18}$\\
    Draco & 4297 & 39472 & 0.111 & 49.8 & -1.0 & $8.2\times10^{-3}$& $2.6\times10^3$ & 1.96 & 6.09 & 0.95 & 1.41 & $1.3\times10^{19}$\\
    \hline\hline
    \end{tabular}
    \caption{Table of the four dwarf spheroidal galaxies considered by VERITAS in this analysis, showing the VERITAS observational results in the first five columns and the assumed properties of the dwarf galaxies. Columns 1 to 5 show the counts recorded by VERITAS in the ON and OFF regions, ratio between the areas of the ON and OFF regions, and the exposure times and detection significances, respectively. The next five columns give the selected parameter set of the generalized NFW profile \citep[described in Section~\ref{sec:method}]{GS2015} for the four dwarf spheroidal galaxies considered. The final two columns show the maximum angular distance considered in the $J$-factor calculation and the $J$-factor, respectively.}
    \label{tab:nfw}
\end{table*}

\section{VERITAS Observatory} \label{sec:veritas}
VERITAS is an array of four Imaging Atmospheric Cherenkov Telescopes (IACTs). The instrument is located at the Fred Lawrence Whipple Observatory in southern Arizona (31$^\circ$40$^\prime$ N 110$^\circ$57$^\prime$ W). The telescope optics utilize a Davies-Cotton design. The reflectors are 12 m in diameter and composed of 350 hexagonal mirrors. The VERITAS cameras are composed of 499 photomultiplier tubes (PMTs), and have a field of view of 3.5$^{\circ}$~\citep{VERITASInstrument}. VERITAS precisely reconstructs $\gamma$-rays with energies between $\sim$100 GeV and $\sim$30 TeV and is sensitive to even higher energy $\gamma$-rays, up to $\sim$100 TeV. This is of particular relevance for this study, corresponding to sensitivity to an annihilation signal from an UHDM particle with mass up to a few tens of PeV. The angular resolution of VERITAS is $\sim$0.1$^{\circ}$ at 1 TeV (68\% containment), while the energy resolution is 15-20\% at 1 TeV. VERITAS can detect a point source with a flux of 1\% of the Crab Nebula flux in $\sim$25 hours of observation~\citep{Park2015}.

Observations of the four dSphs considered here were made between 2007 and 2013. During this time period, VERITAS underwent two upgrades. The first took place in the summer of 2009, in which the position of one of the telescopes was altered to produce a more symmetric array. The second upgrade was made in the summer of 2012, in which the camera PMTs were exchanged for a model with a higher quantum efficiency and the trigger system was upgraded, yielding a 50\% increase to the photon collection efficiency~\citep{Kieda2013}. As the sensitivity of the instrument and the value of the energy threshold changed with each of these upgrades, dedicated Monte Carlo models and instrument response functions (IRFs; including effective areas, energy dispersion matrices, and point spread functions) are available for each of the three array epochs. All data were collected in wobble mode~\citep{Fomin1994}.

Data were reduced using one of the standard VERITAS calibration and event reconstruction pipelines~\citep{VEGAS}. As described in \cite{dm_veritas}, a novel crescent-background technique was used to define the OFF region for background estimation, while the ON region was centered on the target location. The number of ON and OFF counts, the ratio $\alpha$ between the size of the ON and OFF regions, and the detection significance (Li \& Ma significance; \citealp{Li1983}), are given in Table~\ref{tab:nfw}. No low-level data reanalysis was performed; the event lists and IRFs from \cite{dm_veritas} were used for this analysis.

\section{Method} \label{sec:method}
In the previous VERITAS dark matter study using dSphs \citep{dm_veritas}, the so-called event-weighting method \citep{GS2015} was exploited to search for a dark matter signature in the observed data. In this work, we rather adopt a commonly used and extensively documented method, maximum likelihood estimation (MLE), and perform an unbinned likelihood analysis. To perform the MLE analysis, we introduce a likelihood function, quantifying the consistency of the observed dSph data ($D$) with a given dark matter model,
\begin{equation}
\begin{aligned}
    \mathcal{L} (\langle\sigma v\rangle; b|D) =& \frac{ \left( N_s + \alpha \, b \right) ^{N_{\rm on} } e^{-(N_s+\alpha b)}}{N_{\rm on}!}\frac{b^{N_{\rm off}}e^{-b}}{N_{\rm off}!} \\
    & \prod^{N_{\rm on}}_{i=1}\frac{N_{s}\mathit{p}_{s}(E_i) + \alpha \, b \, \mathit{p}_{b}(E_i)}{N_{s}+\alpha \, b}.
\end{aligned}
\label{eq:lh}
\end{equation}
This likelihood is a product of the likelihoods modeling the total counts in the ON and OFF regions, as well as the predicted energy distribution of the counts in the ON region. In more detail, $N_{\rm on}$ and $N_{\rm off}$ are the number of observed ON- and OFF-region counts, respectively, and $\alpha$ is the relative exposure time between the ON and OFF regions. The nuisance parameter $b$ is the expected number of background counts. Two probability density functions (PDFs) are required in this unbinned likelihood function: one for the dark matter signal ($p_{s}$) and the other for the background ($p_{b}$). The background PDF is obtained from the normalized OFF-region event distribution. 
The dark matter signal PDF and the dark matter signal counts ($N_s$) expected to be observed by the instrument within the ON region, of size $\Delta \Omega$, are determined by the dark matter spectrum ($dN_{\gamma}(E)/dE$) and $J$-factor ($J(\Delta \Omega)$), which is the square of the dark matter density integrated along the line of sight within the ON region. In detail,
\begin{equation}\label{eq:dm_signal}
\begin{aligned}
\frac{dN_s}{dE} = \frac{\langle\sigma v\rangle}{8\pi M_{\chi}^2} \frac{dN_{\gamma}}{dE} J(\Delta \Omega).
\end{aligned}
\end{equation}
Here $\langle\sigma v\rangle$ and $M_{\chi}$ are the velocity-averaged dark matter annihilation cross section and dark matter particle mass, respectively. Although not shown here, these results are convolved with the IRF of VERITAS to obtain $p_{s}$ and $N_s$, which accounts for the finite angular and energy resolution of the instrument. For more details, see \cite{dm_veritas}.

For the $\gamma$-ray spectrum from dark matter annihilation at production, $dN_{\gamma}(E)/dE$, we use \texttt{HDMSpectra}  \citep{HDM}\footnote{\url{https://github.com/nickrodd/HDMSpectra}} instead of the widely used \texttt{PPPC4DMID} spectrum \citep{PPPC}. This is because the former provides dark matter annihilation spectra for various channels in a broad mass range from 1 TeV up to the Planck energy, while \texttt{PPPC4DMID} extends only to a dark matter mass of 100 TeV. With \texttt{HDMSpectra}, we obtain a set of nine final state photon spectra, assuming a 100\% branching ratio of dark matter particles in nine different annihilation channels: $e^{+}e^{-}$, $\mu^{+}\mu^{-}$, $\tau^{+}\tau^{-}$, $t\bar{t}$, $b\bar{b}$, $W^{+}W^{-}$, $ZZ$, $\gamma\gamma$, and $\nu_e \bar{\nu}_e$. In considering the differences between the production spectrum and the photon spectrum observable by VERITAS, it is important to note that the UHDM signature (from e.g., the annihilation of a 30 PeV dark matter particle) results in observed $\gamma$-rays below 100 TeV. Consequently, absorption on ambient photon fields can be ignored.

For the dark matter density profile, $\rho(r)$, we adopt the generalized Navarro-Frenk-White (NFW) profile \citep{Hernquist1990, Zhao1996, GS2015}, which is a function of five parameters, 
\begin{equation} \label{eq:dm_profile}
    \rho(r) = \frac{\rho_{s}}{(r/r_s)^{\gamma}[1+(r/r_s)^{\alpha}]^{(\beta-\gamma)/\alpha}}.
\end{equation}
The values of the free parameters used for each dSph and the resulting (unconvolved) $J$-factors are given in Table~\ref{tab:nfw}. The parameters are adopted from \cite{GS2015}.

For the joint-fit analysis, in which data from the four dSphs are combined to maximize statistical power, we combine the individual likelihood functions to form a joint one,
\begin{equation}
    \mathcal{L}_{\rm joint}(\langle\sigma v\rangle; \mathbf{b}|\mathbf{\mathcal{D}}) = \prod_{i = 1}^{N_{\text{target}}} \mathcal{L}(\langle\sigma v\rangle; b_i|D_i).
\end{equation}

The significance of the dark matter signal over background can be obtained by comparing two likelihoods, 
\begin{equation}
    \lambda = -2 \ln \left(\frac{\mathcal{L}_{N_s \equiv 0}}{\mathcal{L}_{N_s \neq 0}} \right)\!.
\end{equation}
If the significance of the dark matter signal is below the threshold to claim a detection ($\lambda \gtrsim$ 25), we compute an upper limit on the dark matter velocity-weighted annihilation cross section by using the likelihood profile. The one-sided 95\% confidence level upper limit on the dark matter velocity-weighted annihilation cross section is the value of the cross section corresponding to $\Delta \ln\mathcal{L}$ of 1.35 compared to the likelihood maximum.

\section{Results} \label{sec:result}

\begin{figure*}[t!]
    \centering
    \subfigure{\includegraphics[width=0.3\linewidth]{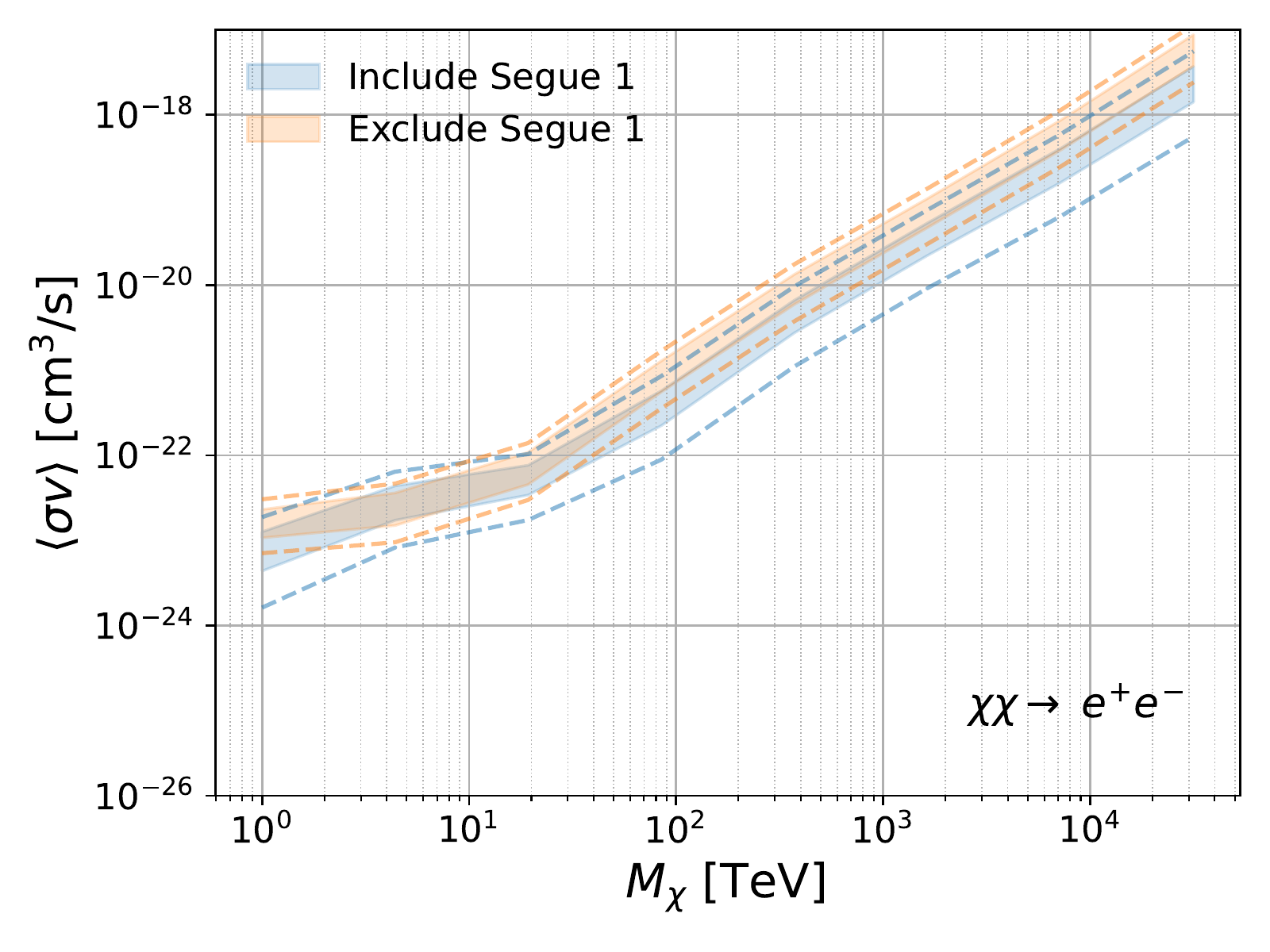}}
    \subfigure{\includegraphics[width=0.3\linewidth]{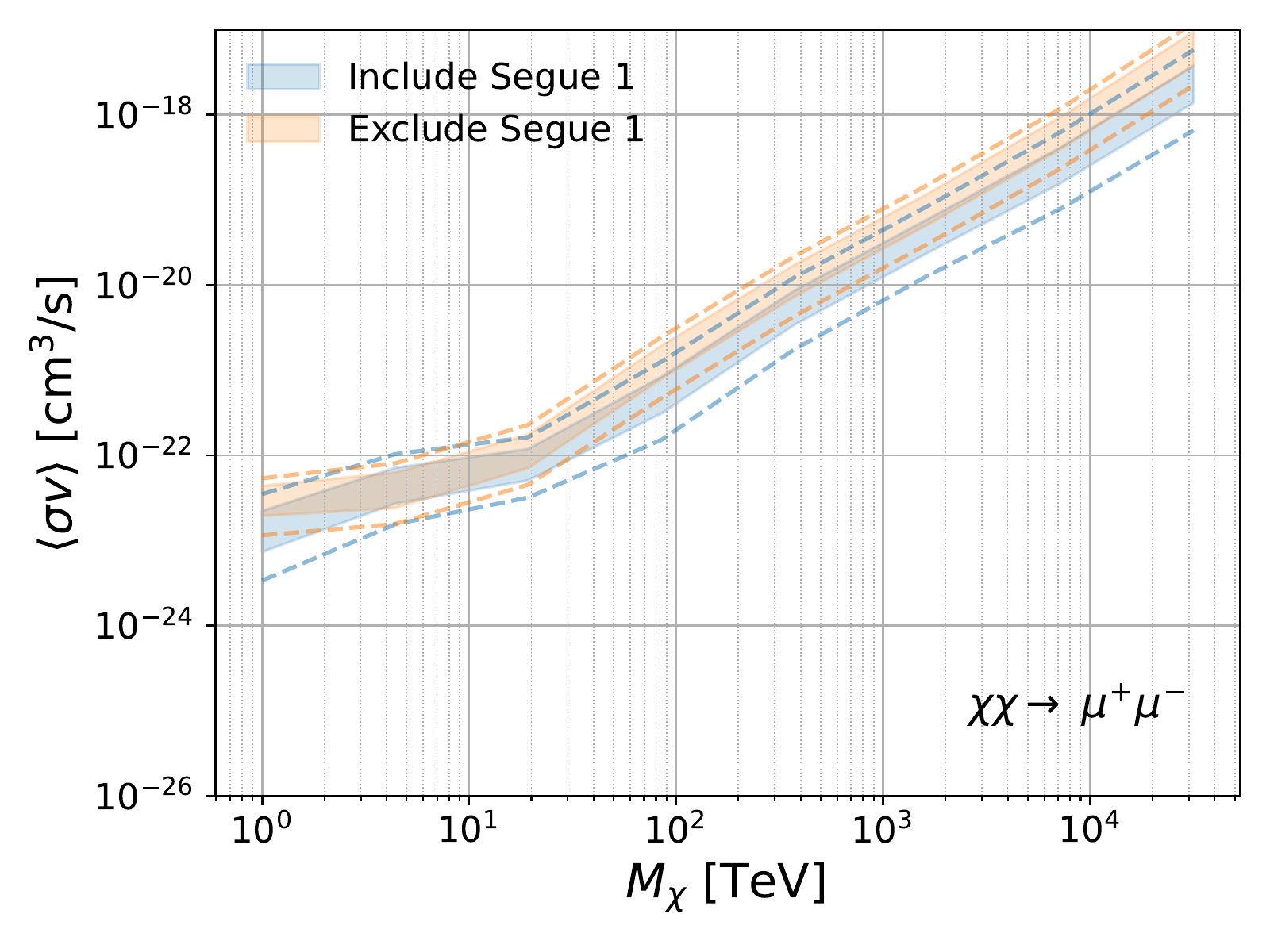}}
    \subfigure{\includegraphics[width=0.3\linewidth]{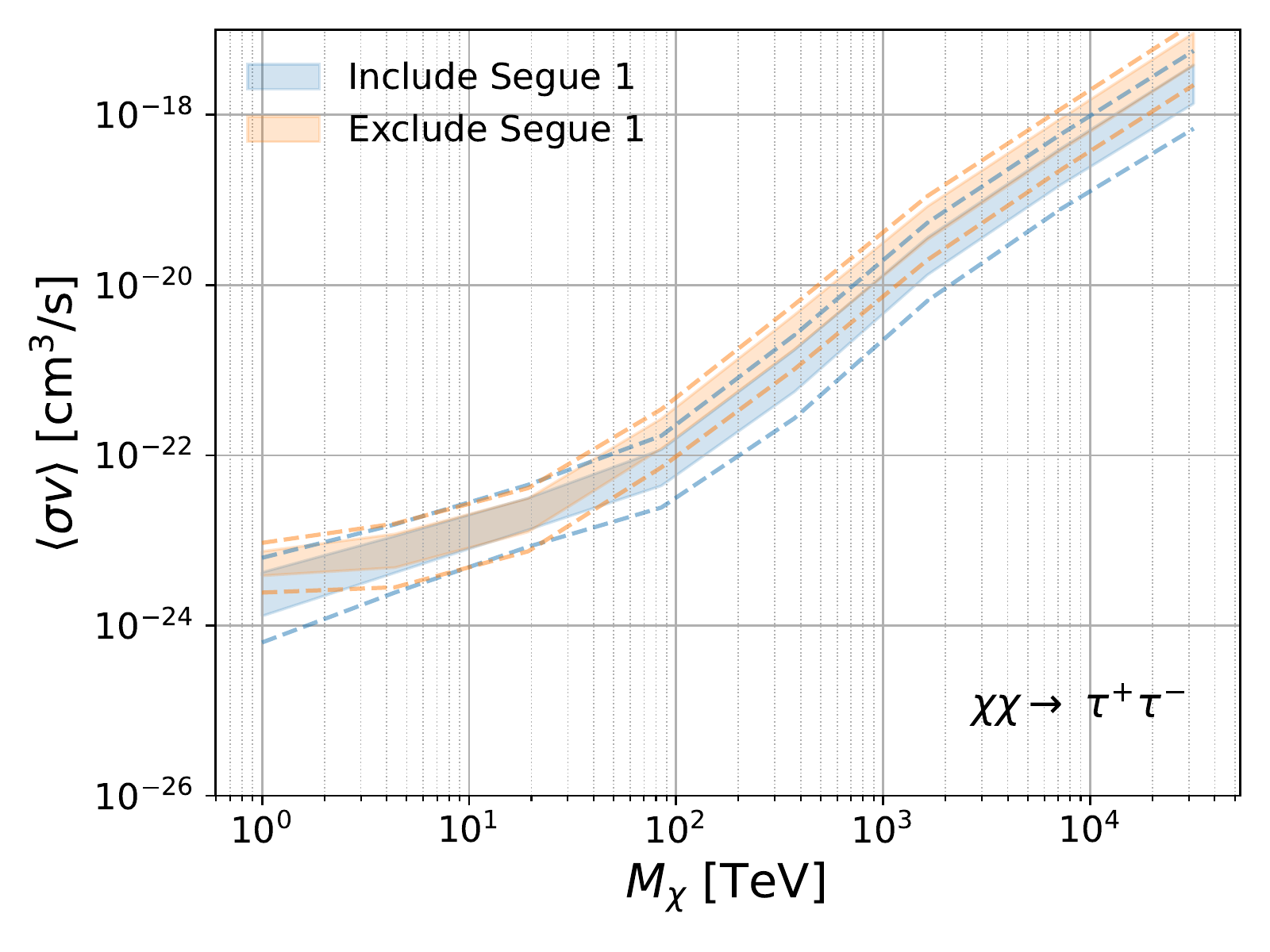}}\\
    \subfigure{\includegraphics[width=0.3\linewidth]{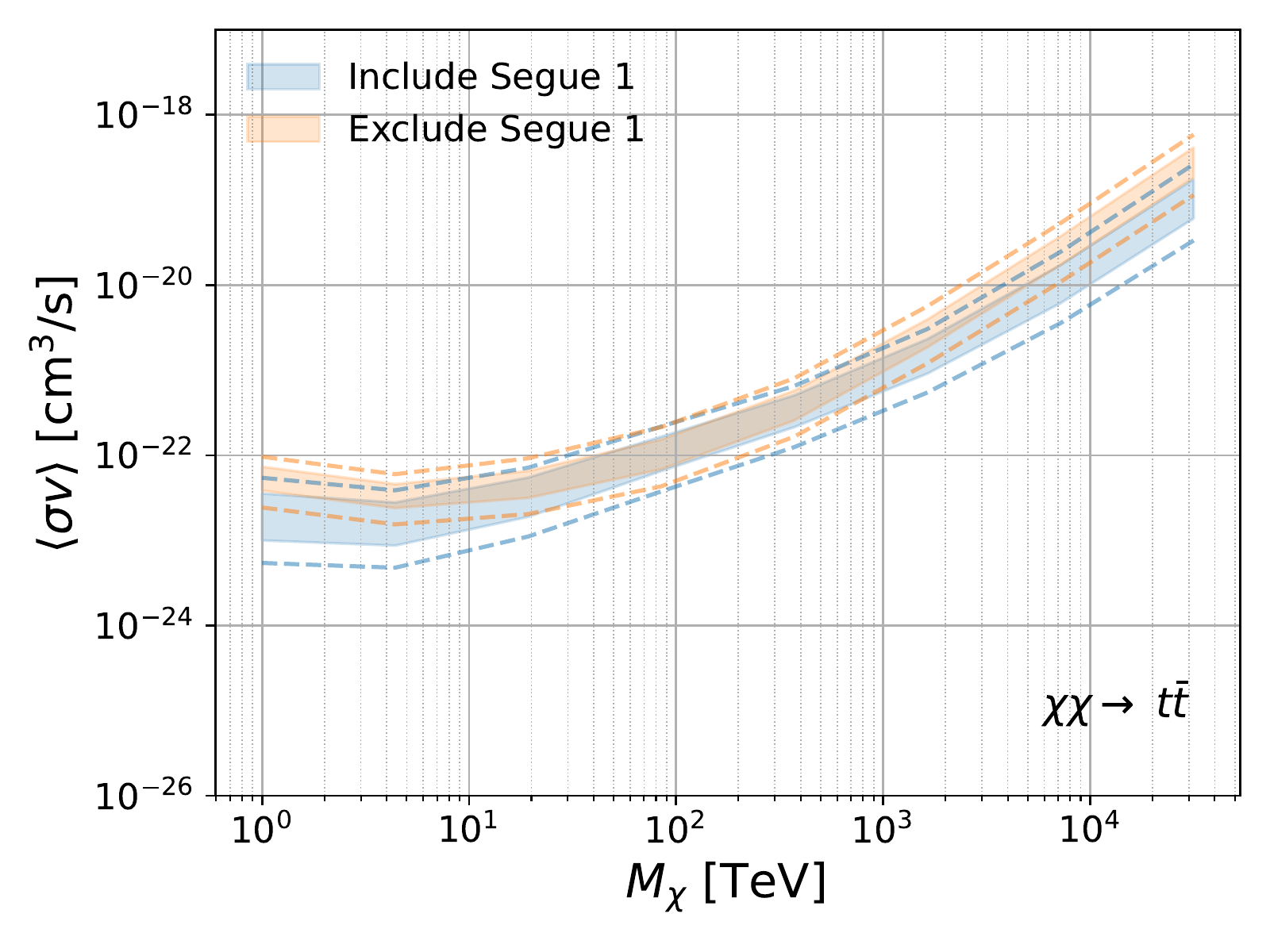}}
    \subfigure{\includegraphics[width=0.3\linewidth]{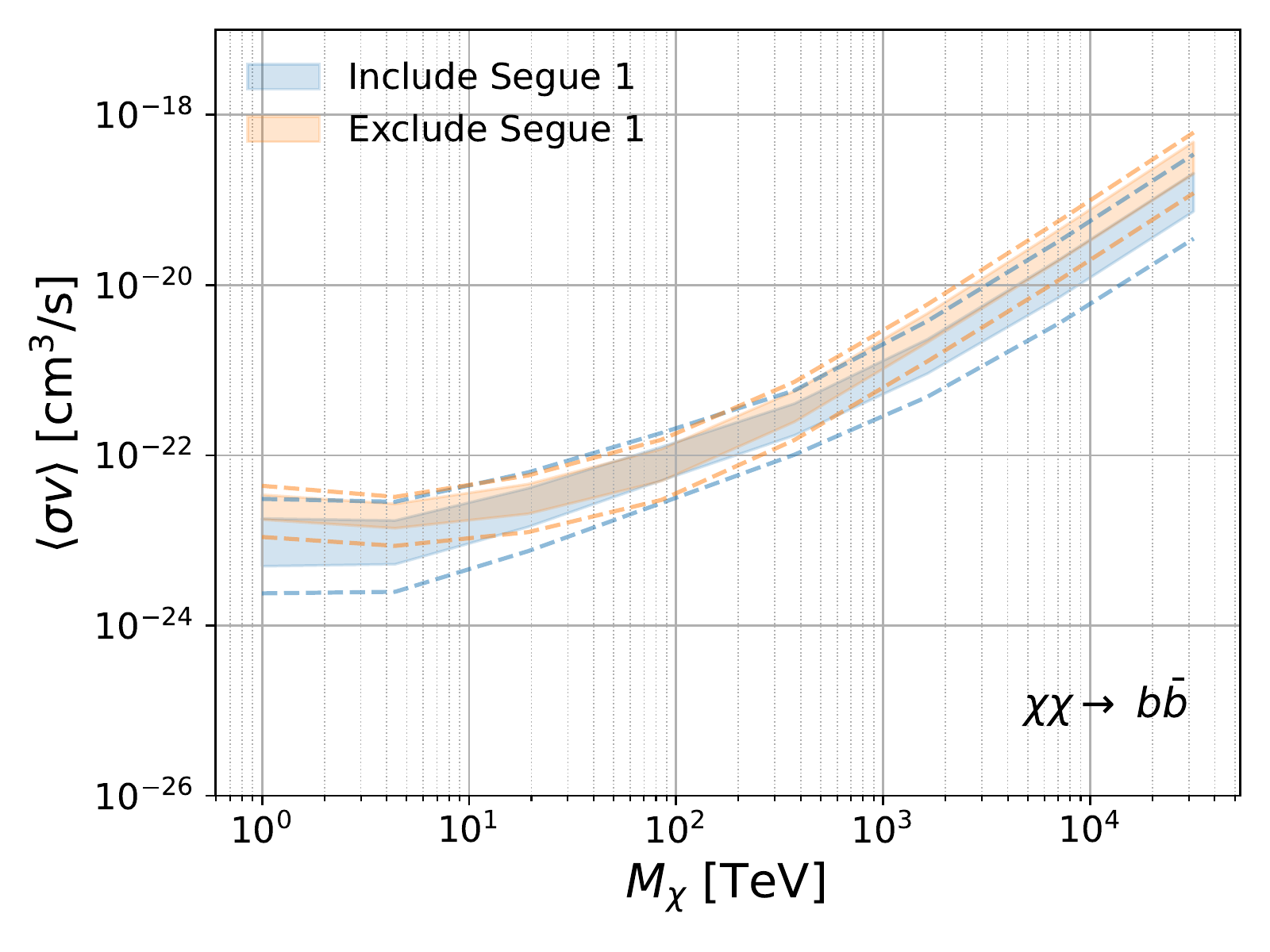}}
    \subfigure{\includegraphics[width=0.3\linewidth]{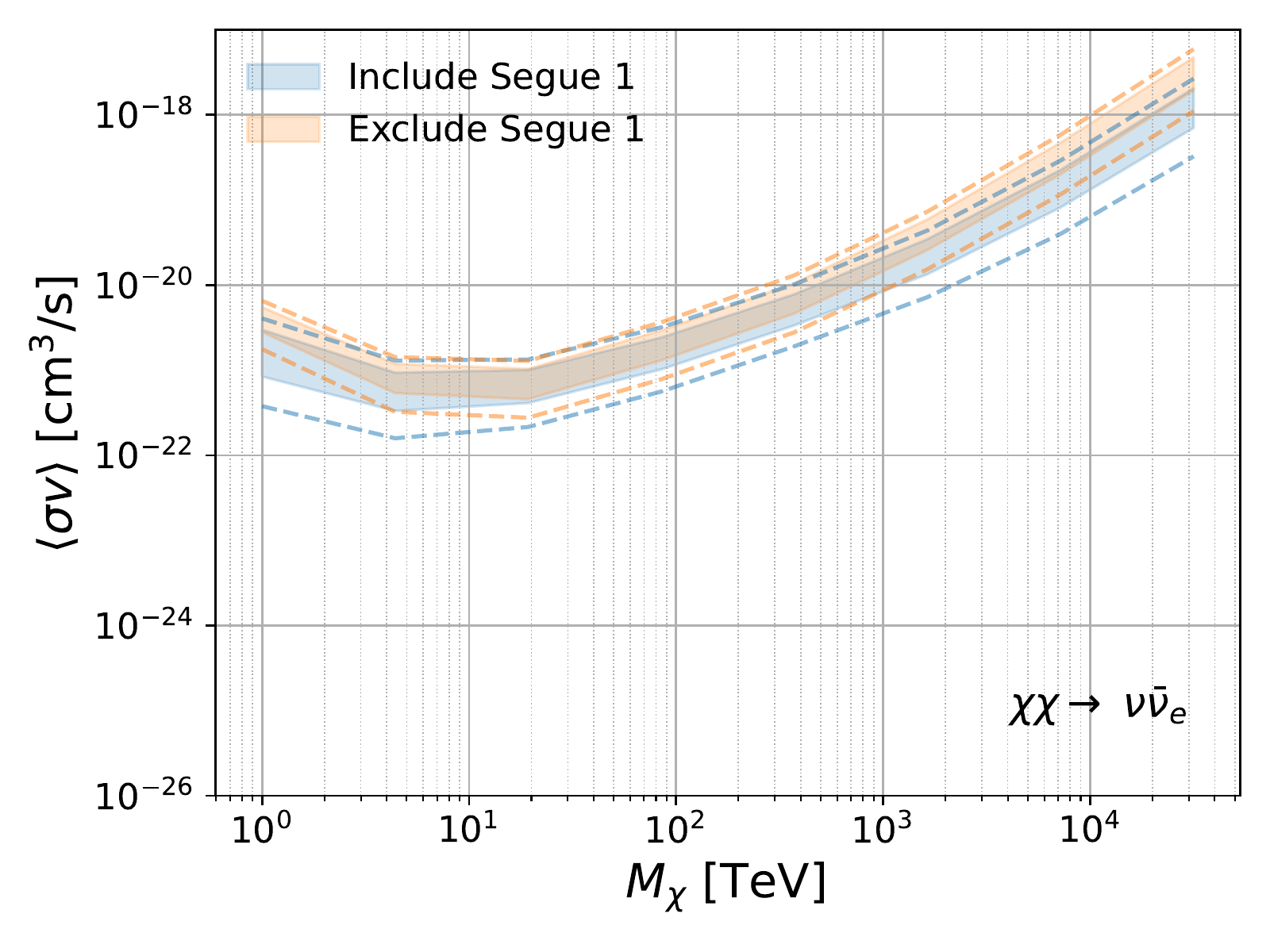}}\\
    \subfigure{\includegraphics[width=0.3\linewidth]{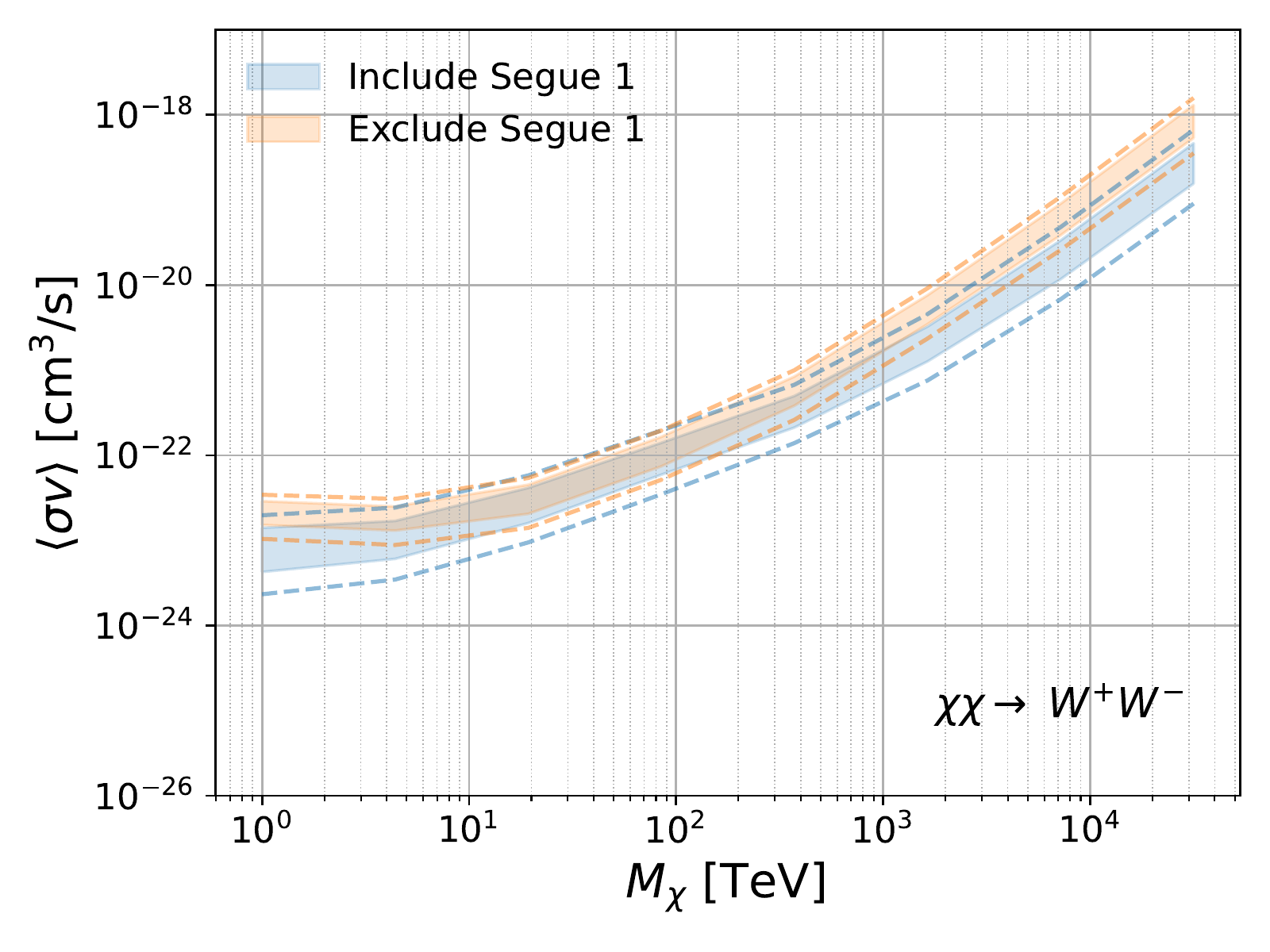}}
    \subfigure{\includegraphics[width=0.3\linewidth]{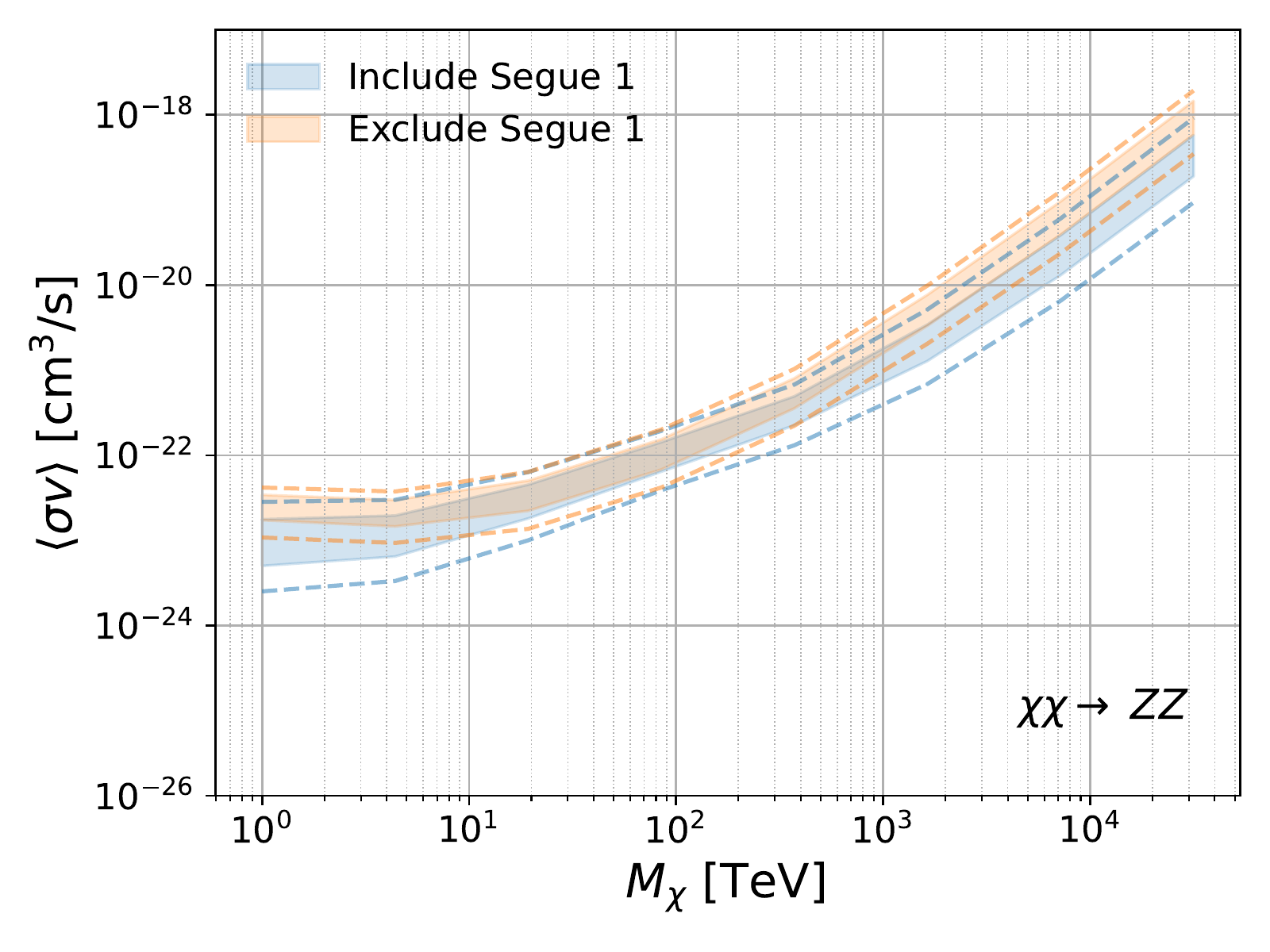}}
    \subfigure{\includegraphics[width=0.3\linewidth]{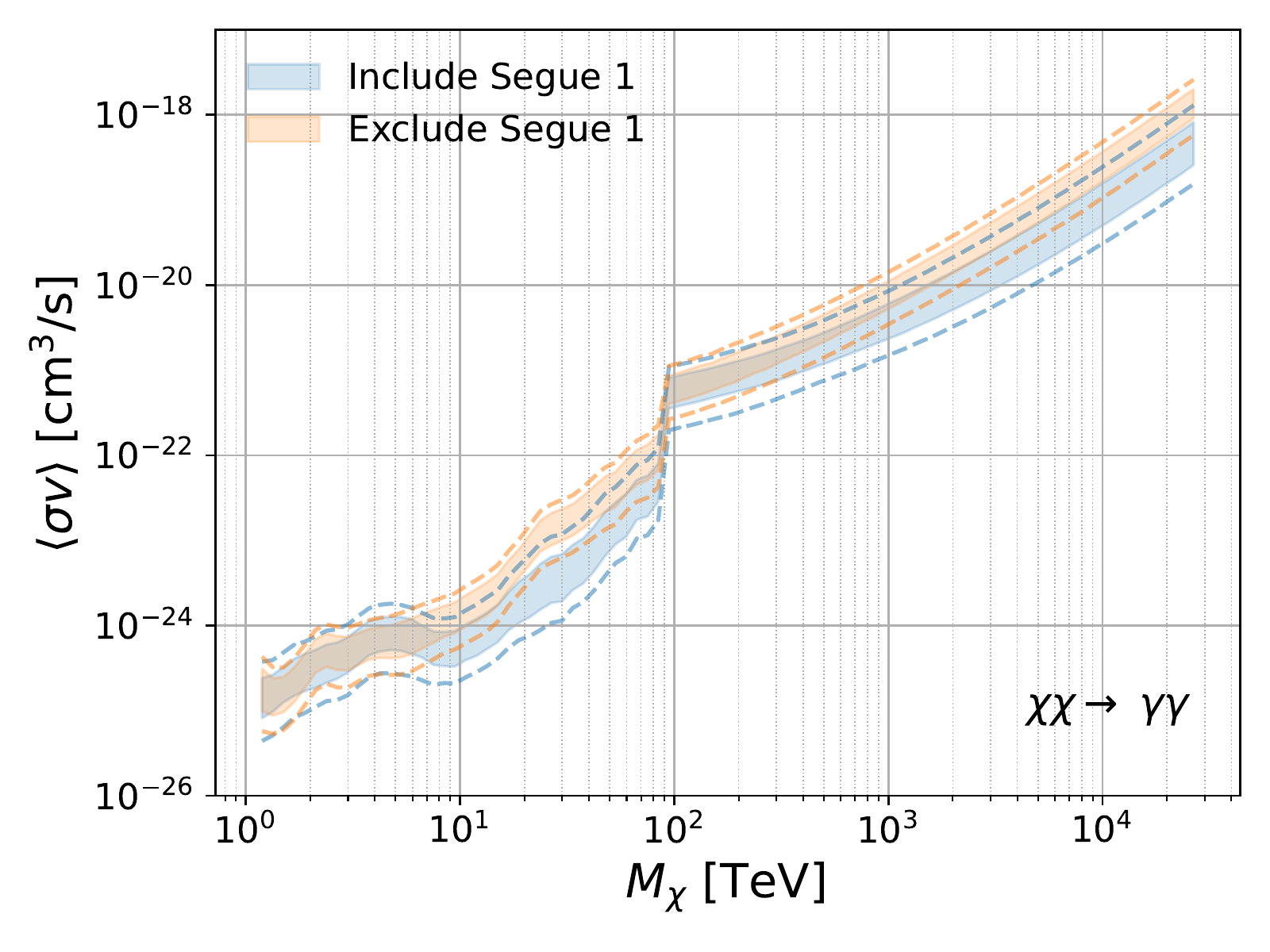}}
    \caption{Velocity-weighted annihilation cross section upper limits produced from VERITAS observations by channel with their systematic uncertainty bands. Due to the uncertainty on the Segue 1 profile, we present upper limits with Segue 1 (blue) and without Segue 1 (orange). A solid (dotted-line) uncertainty band depicts the a 68\% (95\%) containment obtained from 300 realizations of viable dark matter density profiles.}
    \label{fig:comb_uls}
\end{figure*}

\begin{figure*}[t!]
    \centering
    \subfigure{\includegraphics[width=0.3\linewidth]{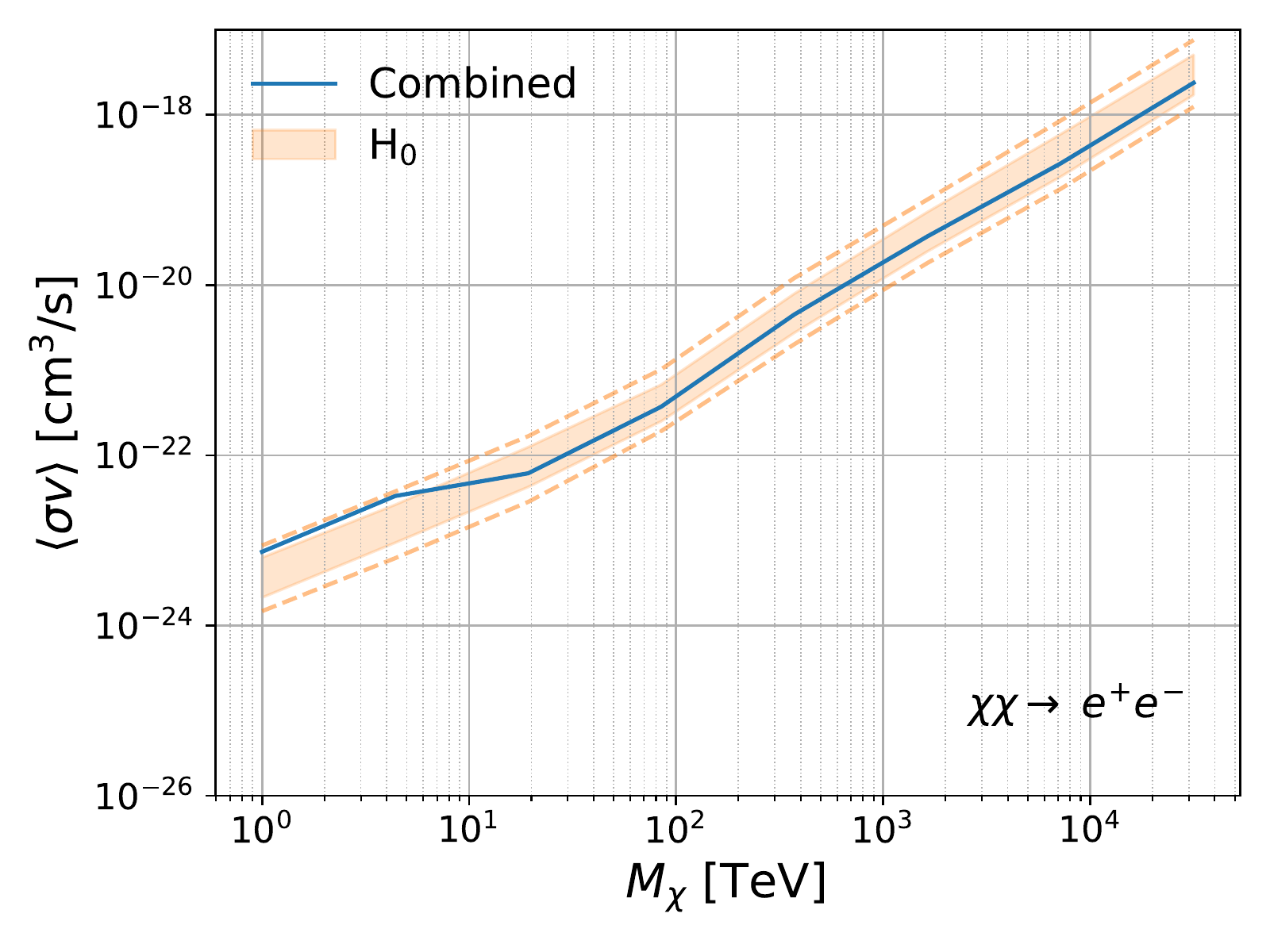}}
    \subfigure{\includegraphics[width=0.3\linewidth]{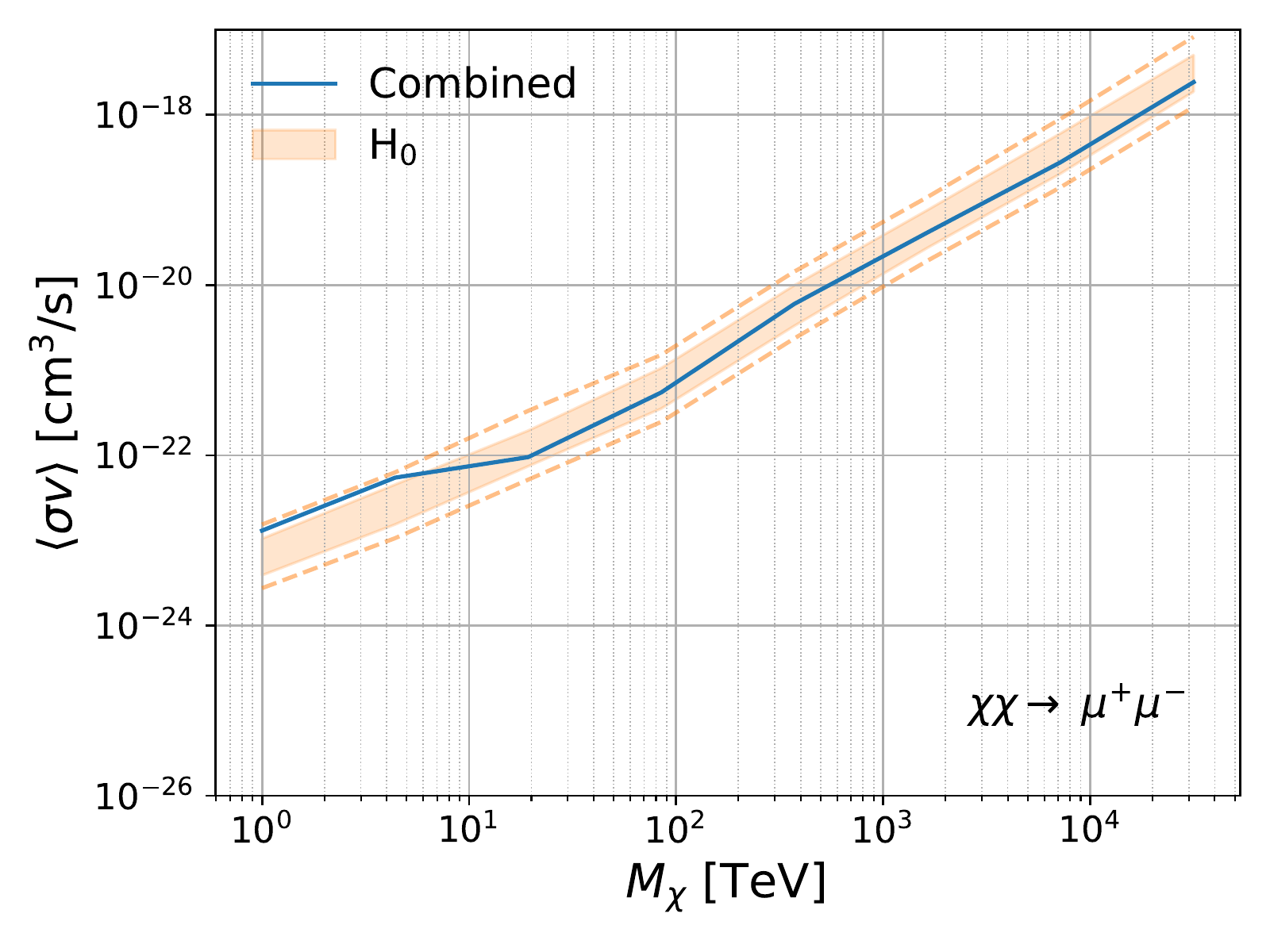}}
    \subfigure{\includegraphics[width=0.3\linewidth]{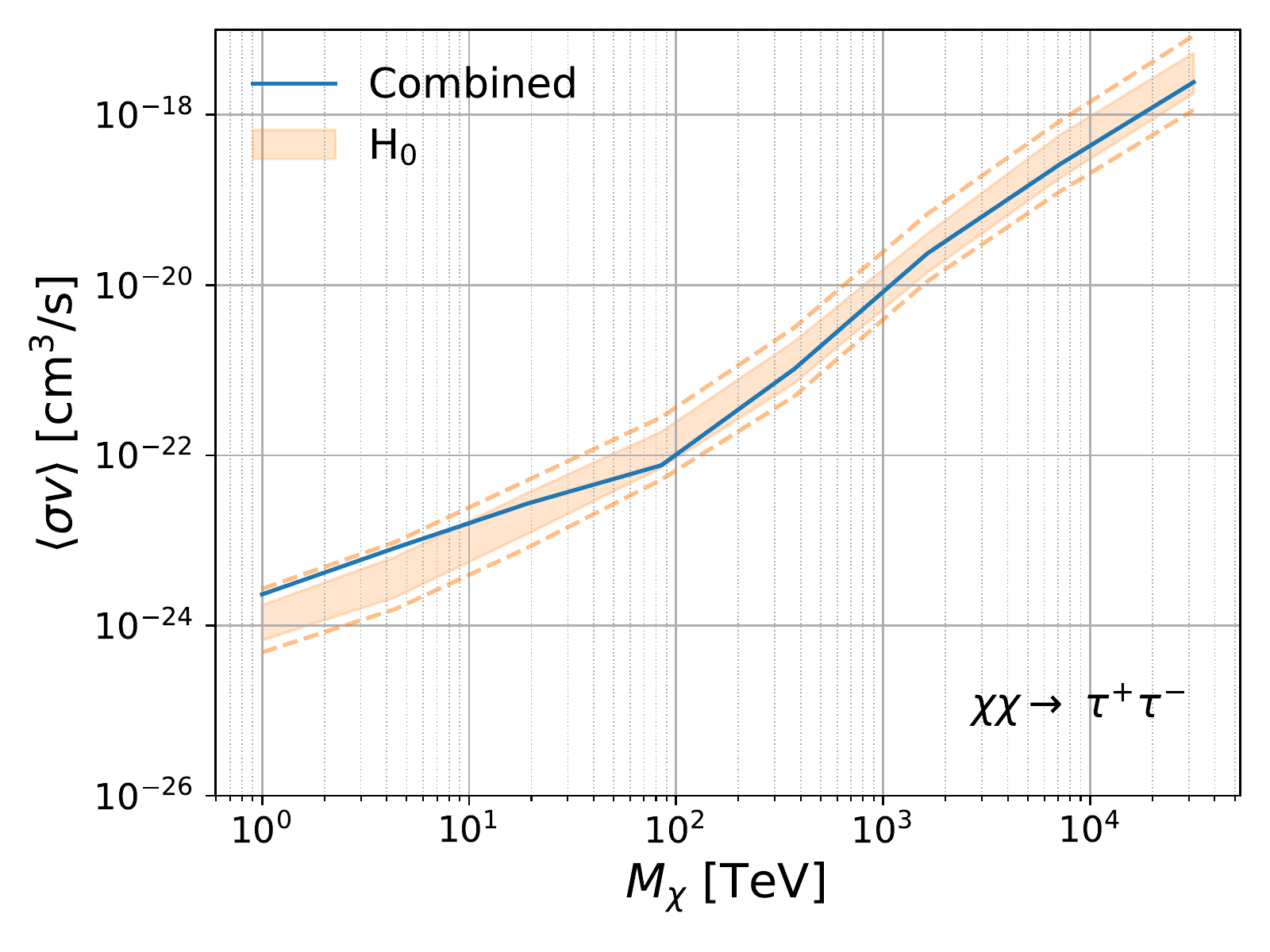}}\\
    \subfigure{\includegraphics[width=0.3\linewidth]{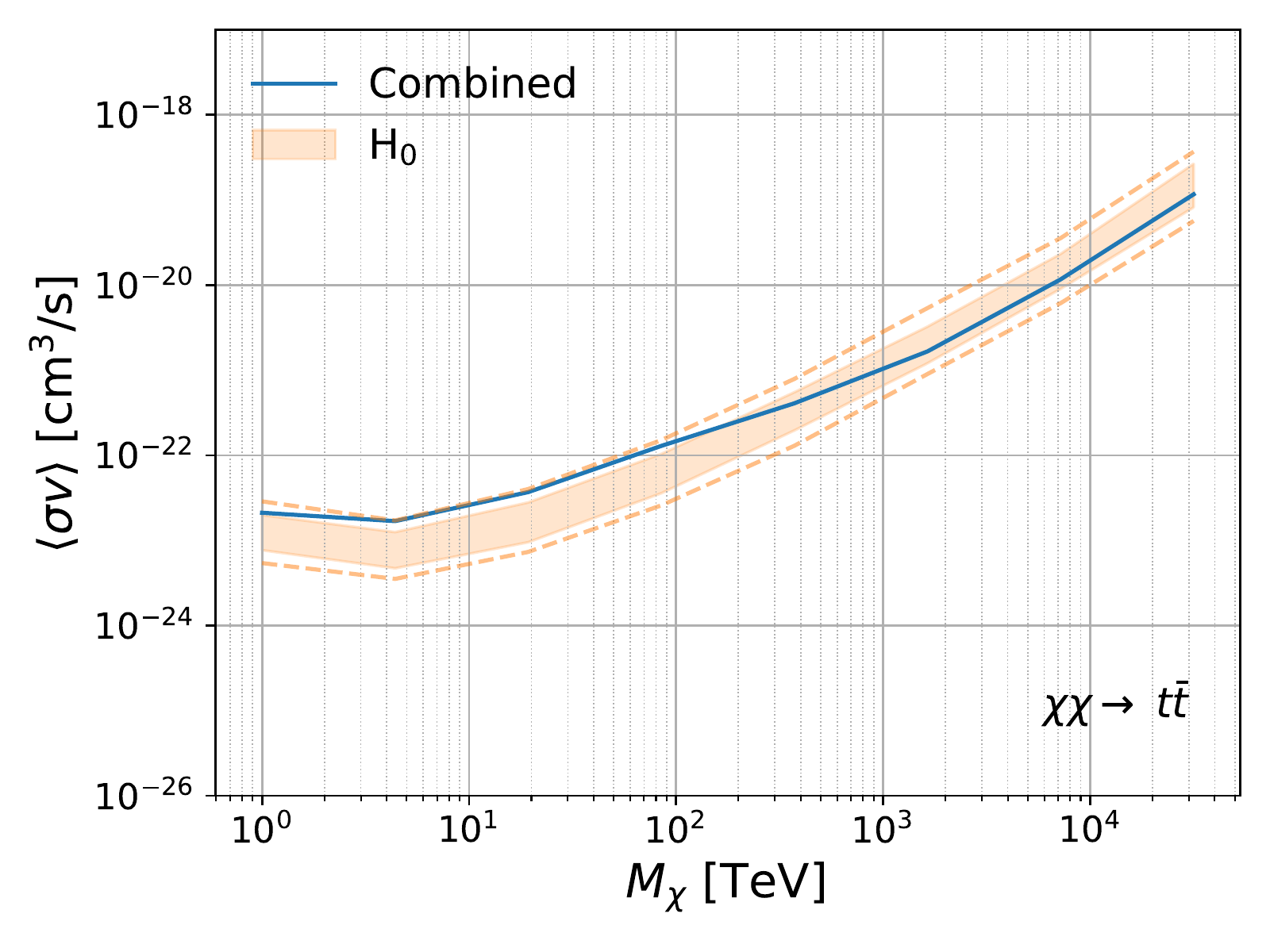}}
    \subfigure{\includegraphics[width=0.3\linewidth]{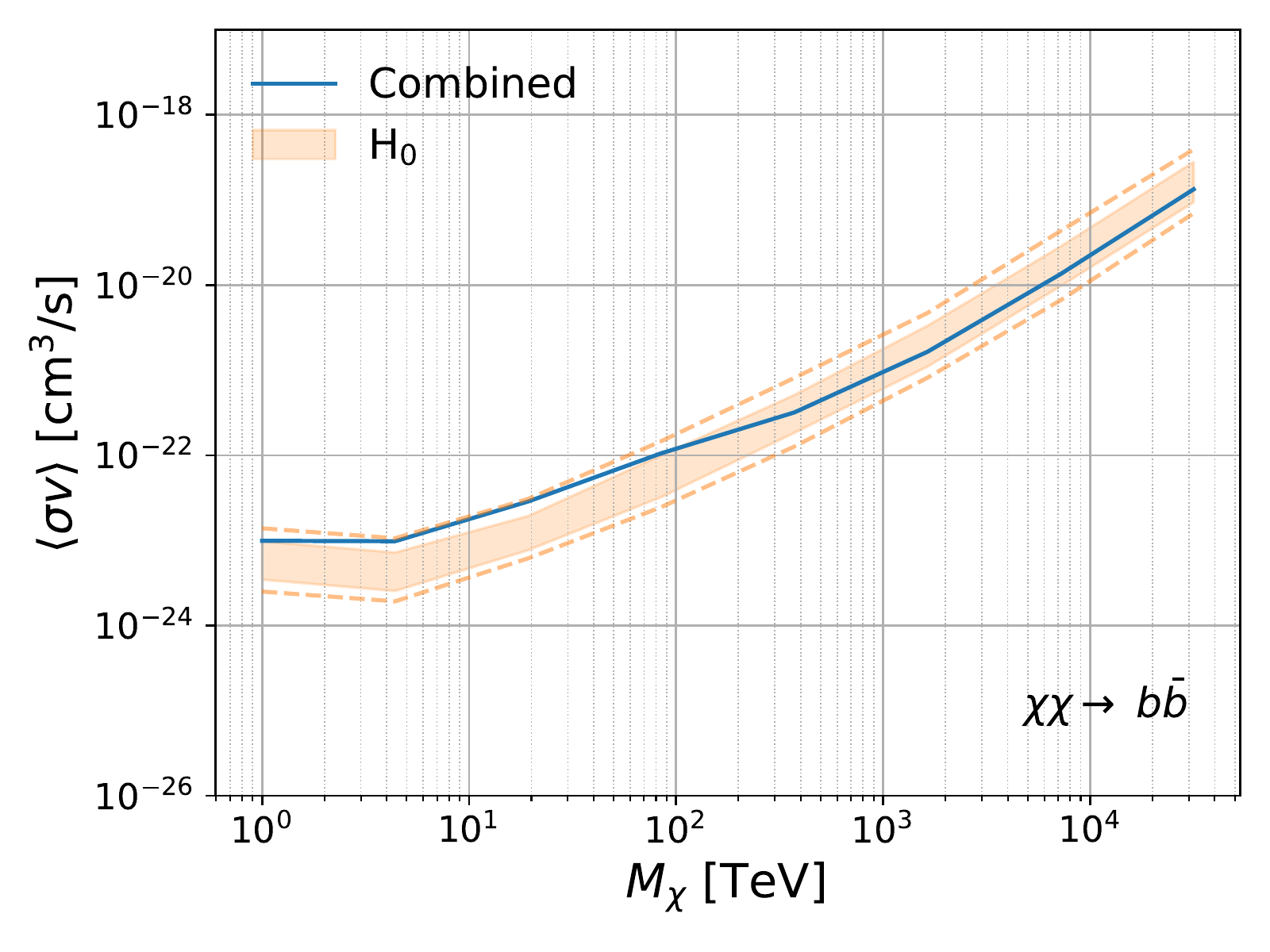}}
    \subfigure{\includegraphics[width=0.3\linewidth]{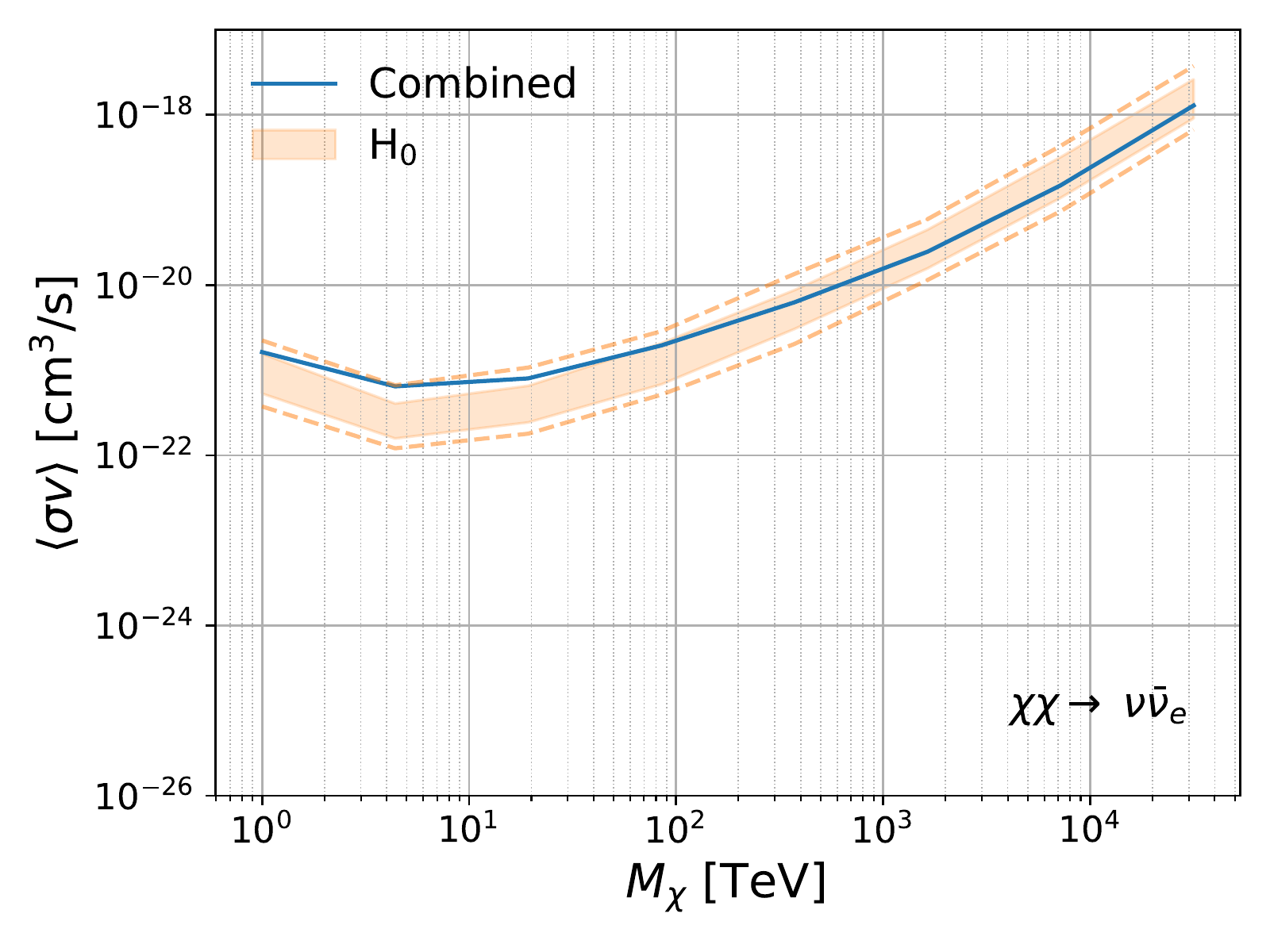}}\\
    \subfigure{\includegraphics[width=0.3\linewidth]{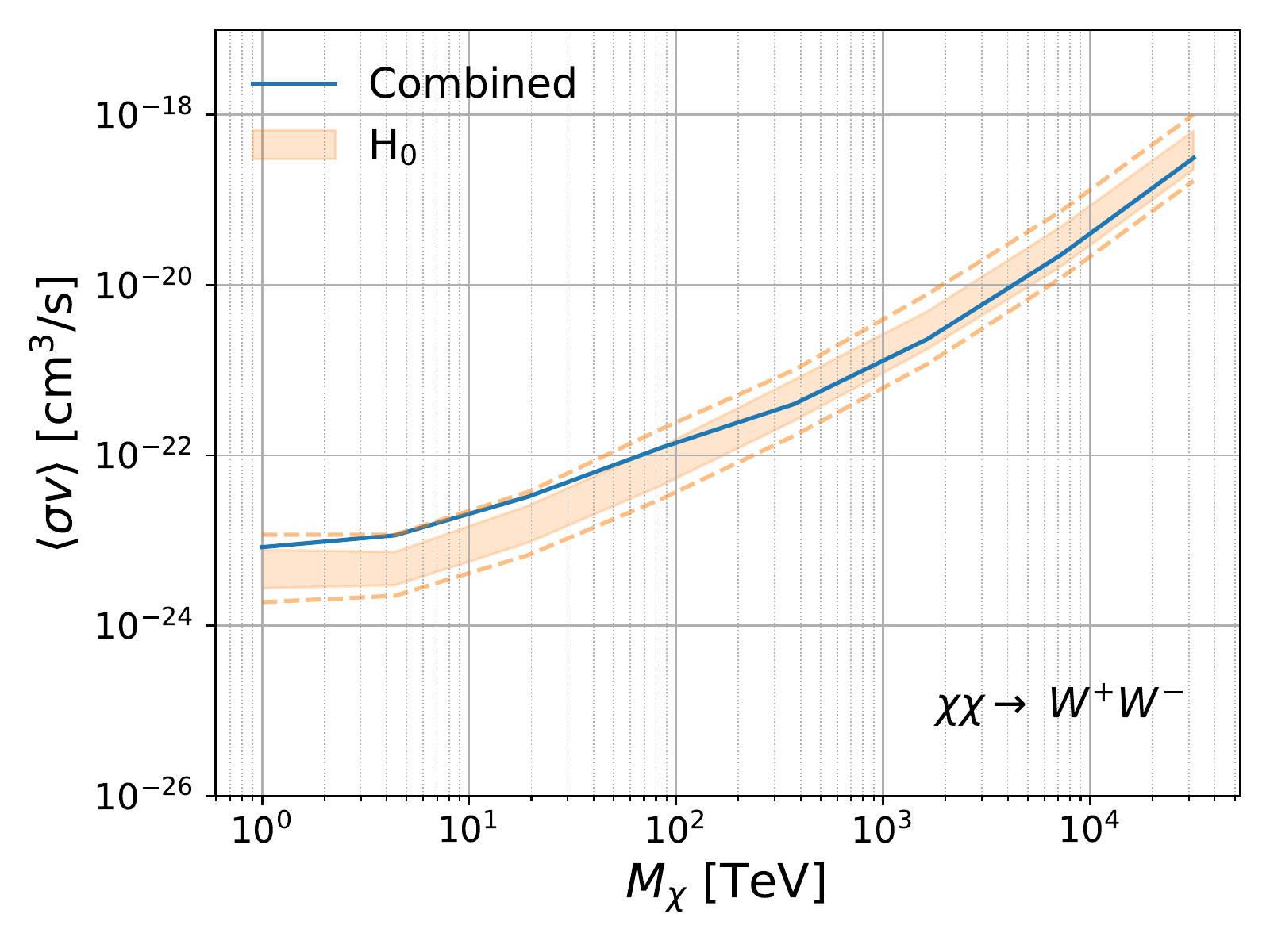}}
    \subfigure{\includegraphics[width=0.3\linewidth]{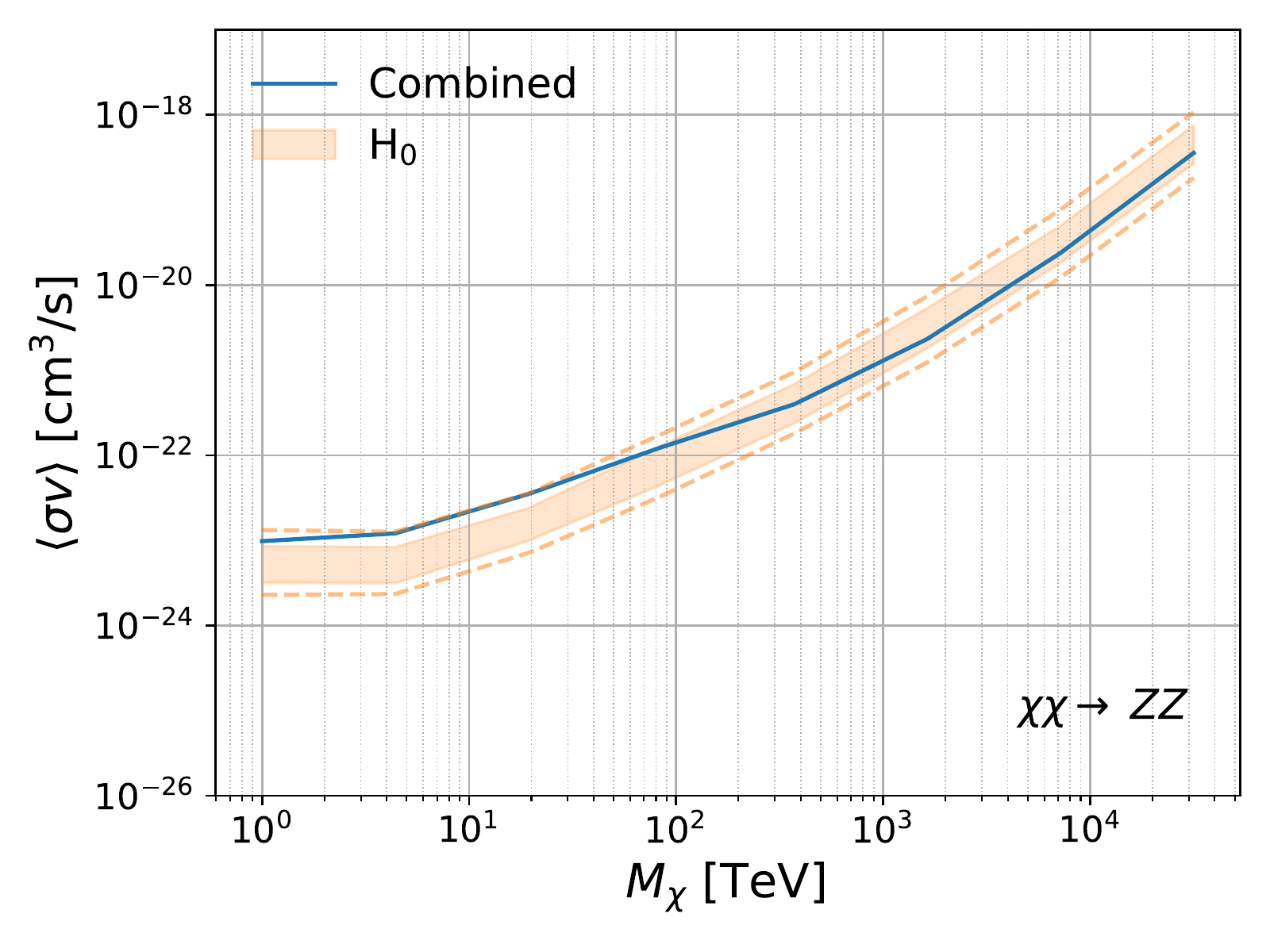}}
    \subfigure{\includegraphics[width=0.3\linewidth]{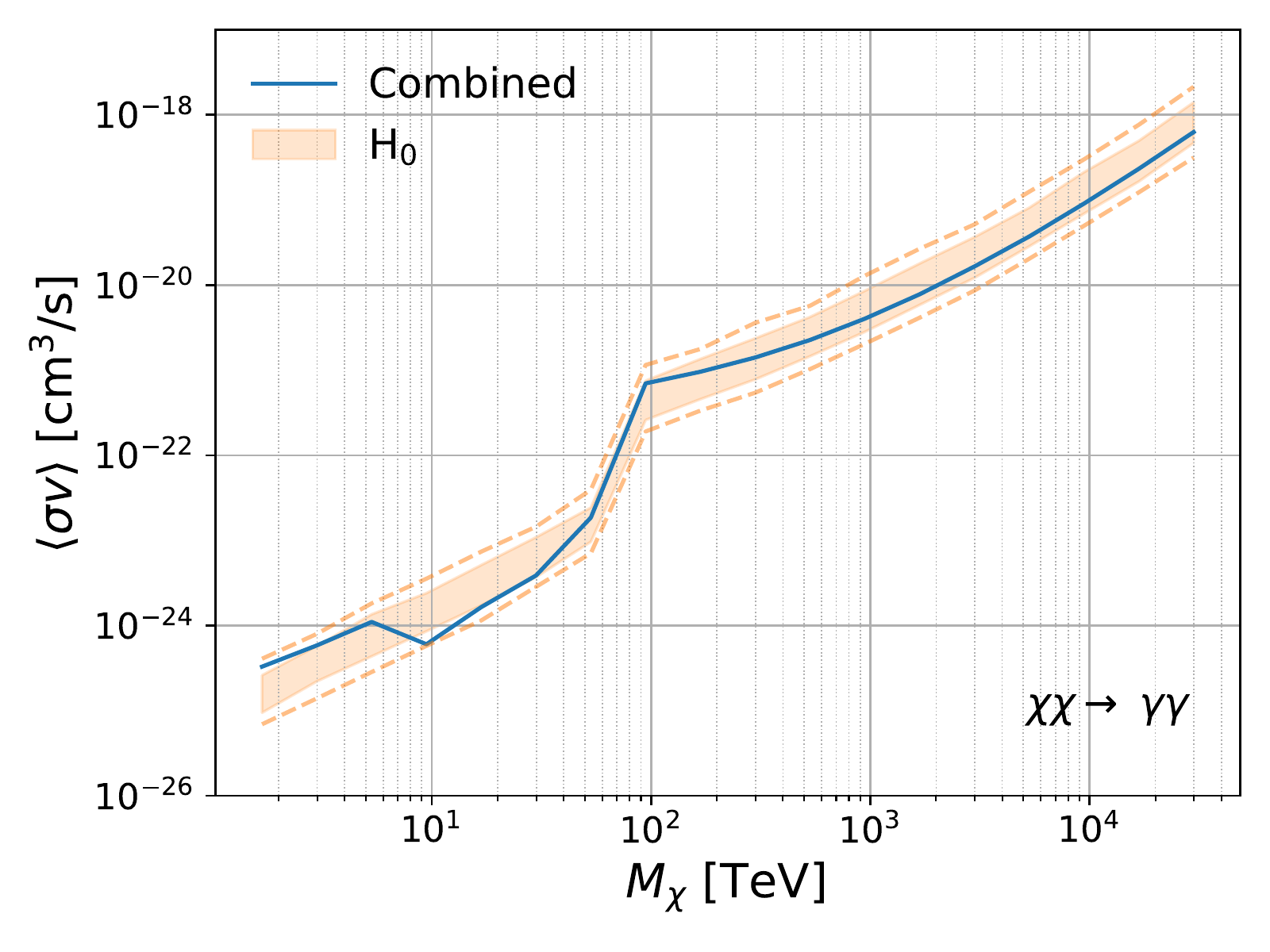}}
    \caption{Velocity-weighted annihilation cross section upper limit curves produced from VERITAS observations by channel compared with their null-hypothesis bands (H$_0$; $\langle\sigma v \rangle = 0$). We present upper limits derived from the four dSph observations (blue) and upper limits with the Poisson background fluctuation (orange). A solid (dotted-line) uncertainty band depicts the 68\% (95\%) containment obtained from 300 realizations of random fluctuations of the background.}
    \label{fig:exp_uls}
\end{figure*}

We do not detect an UHDM signal above background. From the individual and the joint-fit analyses, we obtain $\lambda$ less than our threshold in all annihilation channels and for all masses from 1 TeV to about 30 PeV (see Appendix~\ref{sec:sig}).  

\subsection{Upper limits on the UHDM velocity-weighted annihilation cross section} \label{sec:uls}
We compute upper limits on the dark matter velocity-weighted annihilation cross section for each channel. Figure~\ref{fig:comb_uls} shows upper limits obtained in the joint-fit analysis, each with a systematic uncertainty band resulting from the limited understanding of the dark matter density distribution. The uncertainty band is obtained from 300 realizations with different dark matter density profile parameter sets from \cite{GS2015}; each parameter set can sufficiently describe the stellar-kinematic data observed from the selected target. In the case of Segue 1, the ambiguity of selecting member stars significantly affects the dark matter profile, such that the total density can differ by two orders of magnitude~\citep{Bonnivard2016}. For this reason, we additionally present the combined upper limits excluding the Segue 1 data. 

We note that the discontinuity in the $\gamma\gamma$ channel at around 100 TeV, the maximum value for which we consider $\gamma$-ray events, is expected. Above 100 TeV, the dominant contribution from the delta function/line annihilation signal at $M_\chi = E_\gamma$ results in final state $\gamma$-rays whose energies are above the VERITAS sensitive energy range, leaving only the continuum spectrum. The continuum spectrum is more challenging to detect in comparison to a line signature, resulting in less sensitive limits when the line component is no longer detectable.

\subsection{Comparison with the background fluctuation} \label{sec:expULs}

We test whether the distribution of ON-region events is consistent with the Poisson fluctuation of the background. To do this, we estimate an expected upper limit from a simulated ON region for which events and their energy are randomly selected from the observed OFF-region events. The number of simulated ON events is selected from a Poisson distribution with mean equal to the observed number of OFF-region events, scaled by the ratio of the areas of the ON and OFF regions, $N_{\rm on,sim} = {\rm Pois}(\alpha N_{\rm off,obs})$.  
For each channel, we repeat this process 300 times and obtain an expected upper limit band with the width determined by the magnitude of the background fluctuation. 

Fig.~\ref{fig:exp_uls} shows the comparison of the observed upper limits with the expected upper limit bands. Each solid line (blue) is an upper limit curve from the parameter set listed in Table~\ref{tab:nfw}, and the expected upper limit band is depicted in orange with 68\% (solid) and 95\% (dotted-line) containment. For all annihilation channels, the observed upper limits are consistent with the expected upper limits within the 95\% confidence level. This result supports the non-detection of the UHDM annihilation signal, as well as quantifying the impact of statistical uncertainty on the derived limits.

\section{Discussion and Conclusions} \label{sec:disc}
\begin{figure}[t!]
    \centering
    \subfigure{\includegraphics[width=0.9\linewidth]{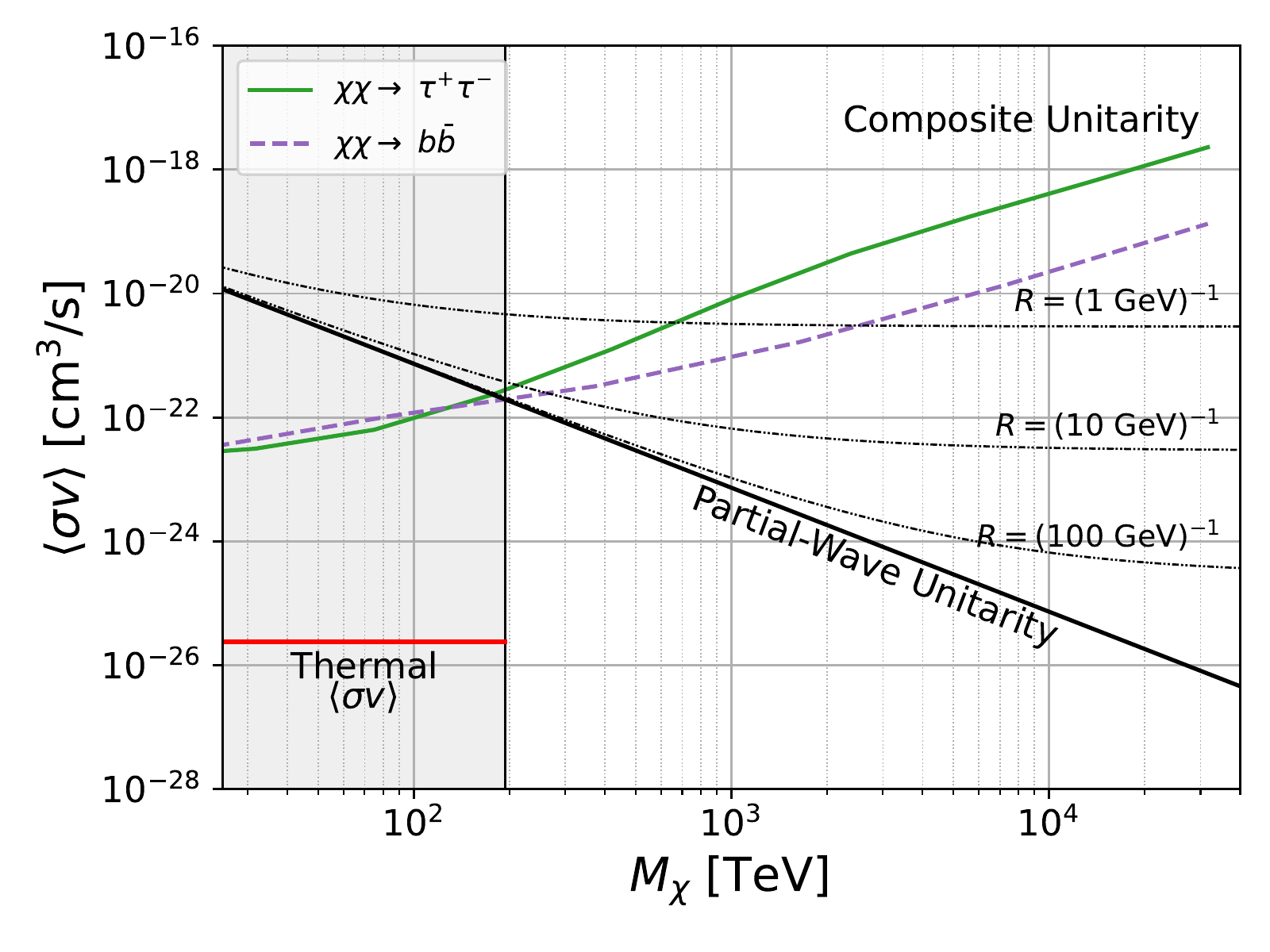}}
    \caption{A comparison of VERITAS upper limit curves for two annihilation channels against UHDM theoretical benchmarks \citep{Tak2022}. The blue solid lines are the 95\% confidence upper limits obtained from the combined analysis and the red solid curve is the thermal-relic cross section ($2.4\times10^{-26}$ cm$^3$/s). The purple line refers to the unitarity limit on a point-like velocity-weighted annihilation cross section for a particle that respects partial-wave unitarity. Above the partial-wave unitarity limit, various composite states can be possible; three possible composite unitary bounds, the purple dashed lines, are plotted as examples.}
    \label{fig:uhdm}
\end{figure}

\begin{figure}[t!]
    \centering
    \subfigure{\includegraphics[width=0.9\linewidth]{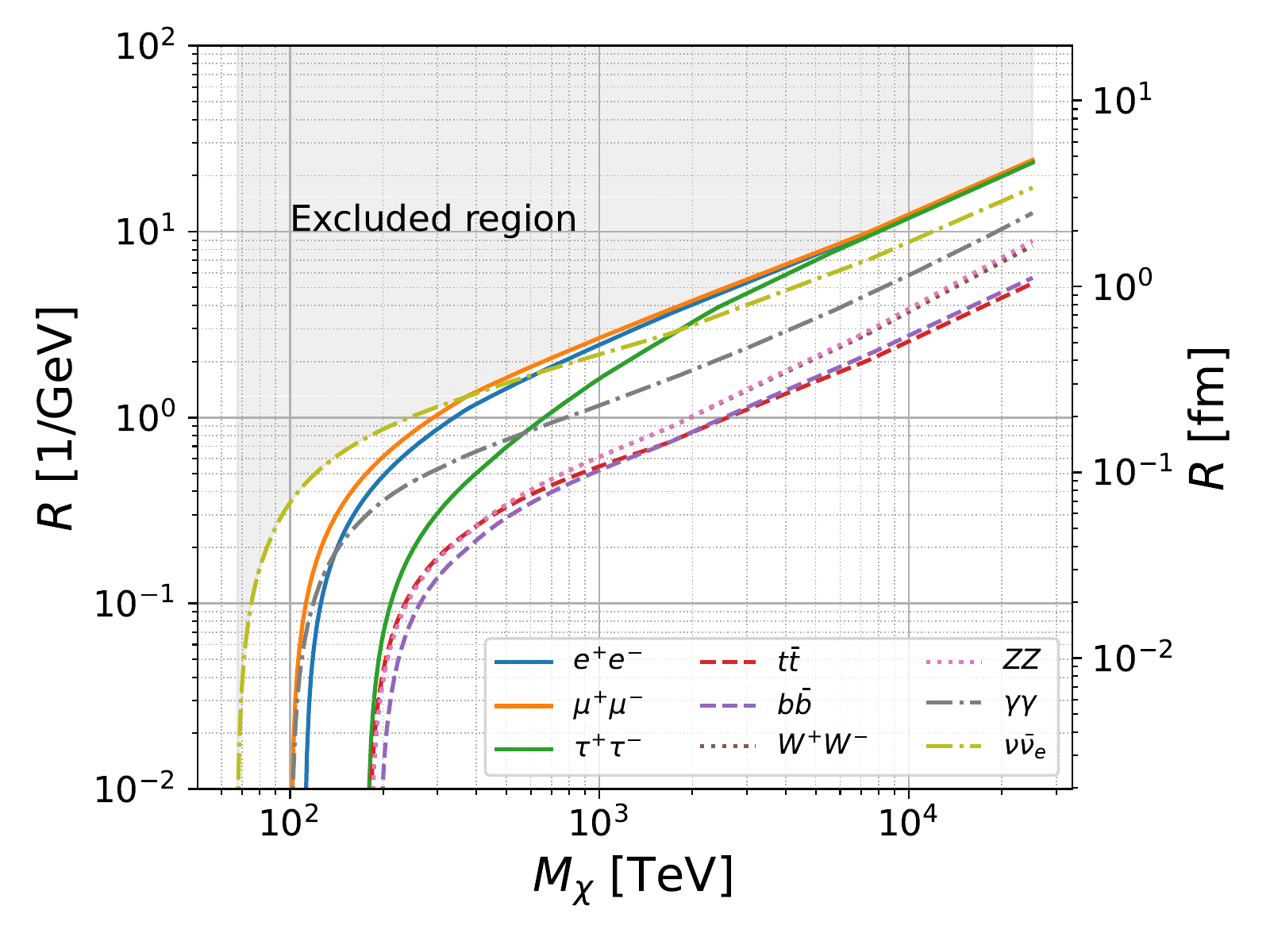}}
    \caption{VERITAS 95\% confidence upper limits curves on the radius, in terms of femtometers and the inverse of energy, of a composite dark matter particle as a function of mass, for the nine annihilation channels considered. The shaded areas denote exclusion regions.}
    \label{fig:uhdmR}
\end{figure}

\begin{figure*}[t!]
    \centering
    \subfigure{\includegraphics[width=0.45\linewidth]{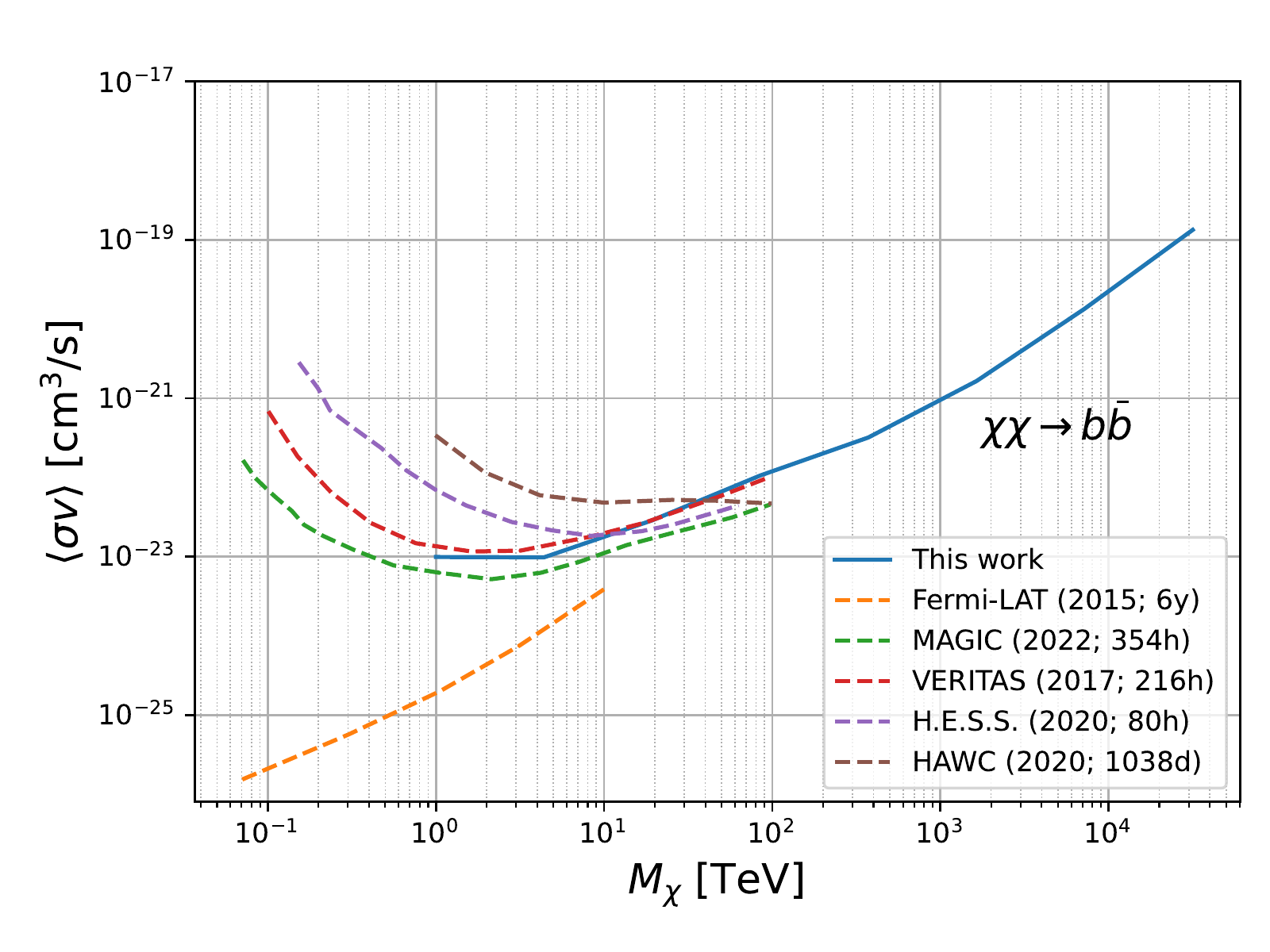}}
    \subfigure{\includegraphics[width=0.45\linewidth]{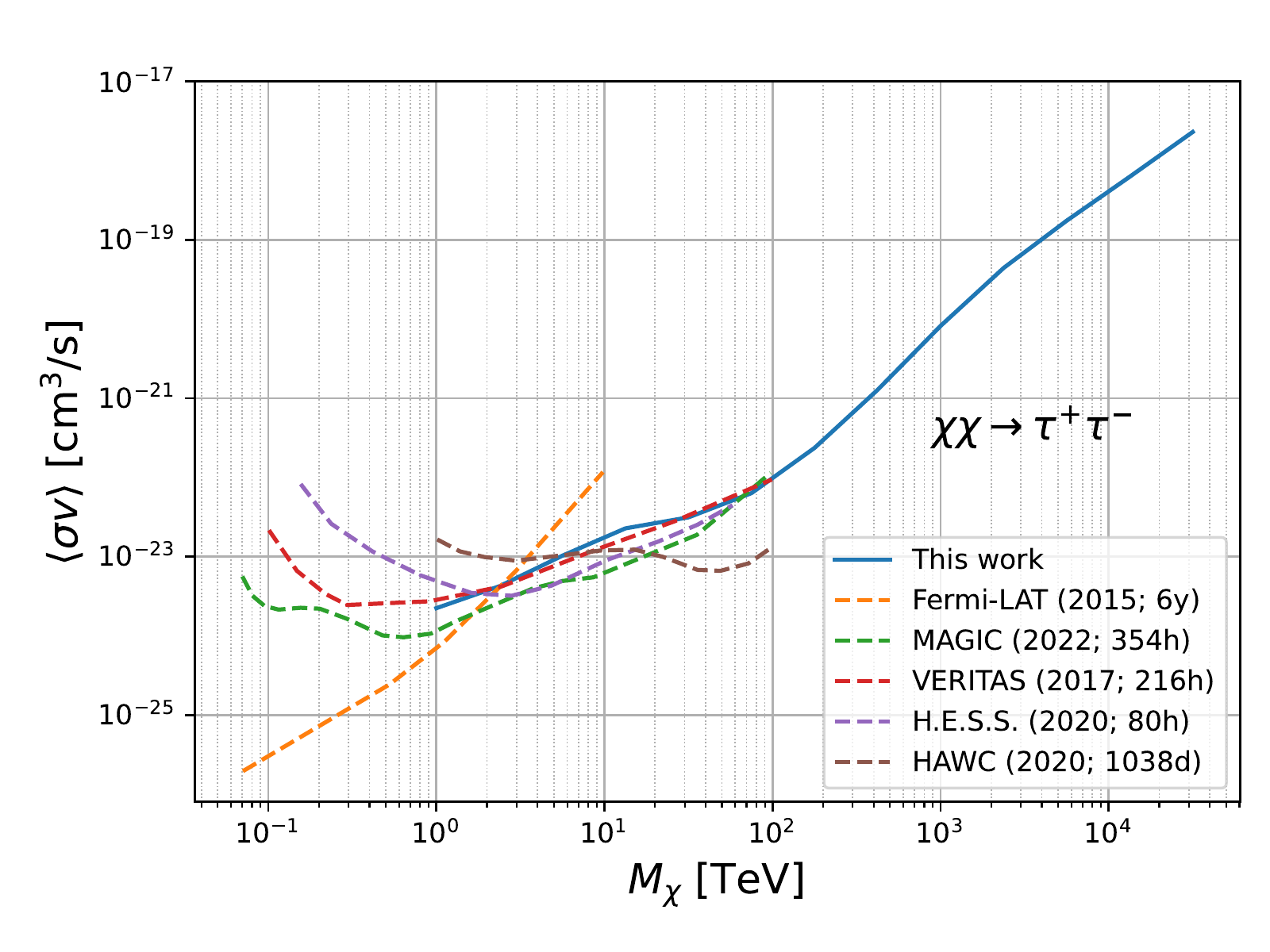}}
    \caption{VERITAS upper limit curves obtained from all four dSphs including Segue 1 compared with other published upper limit curves. All curves show 95\% confidence upper limits on the dark matter velocity-weighted annihilation cross section for the $b\bar{b}$ 
    (left) and $\tau^{+}\tau^{-}$ (right) annihilation channels. They are adapted from \cite{dm_fermi} (Fermi-LAT; orange dashed line), \cite{dm_veritas} (VERITAS; red dashed line), \cite{dm_magic2} (MAGIC; green dashed line), \cite{dm_hawc2} (HAWC; brown dashed line), and \cite{dm_hess2} (H.E.S.S.; purple dashed line). }
    \label{fig:comparison}
\end{figure*}

As mentioned at the outset, a requirement for UHDM is to evade the so-called unitarity limit. The $\mathcal{O}$(100 TeV) bound assumes that the dark matter is a point-like particle, which is in thermal equilibrium with SM particles in the early universe. However, one straightforward way to evade this limit takes point-like dark matter which captures into bound states.  These additional channels can achieve a larger annihilation cross section while respecting unitarity. Individual partial-wave contributions must respect their associated unitarity bound, but the total cross section is given by the sum of all partial-wave contributions. This effect is even seen in medium-sized representations of electroweak SU(2), allowing them to be simple thermal-relic UHDM candidates \citep{Bottaro:2021snn}.

One class of UHDM models that further relaxes the unitarity bounds on mass are composite dark matter models, where UHDM is not a point-like particle and thus possibly has a geometrical cross section. In the case of an interaction with a geometrical cross section, the unitarity bound becomes
\begin{equation}
\langle \sigma v \rangle \leq 4\pi\frac{(1+M_{\chi} v_{\rm rel} R)^2}{(M_{\chi}^2 v_{\rm rel})},
\end{equation}
where $v_{\rm rel}$ is the average velocity between dark matter particles (in our case, in dSph halos), and $R$ is the size of the particle. Note that the unitarity limit for a point-like particle can be reproduced with $R$ = 0, whereas if the particle mass is large enough, we can reproduce the classical cross section of $\langle \sigma v \rangle = 4\pi R^2 v_{\rm rel}$. For a detailed discussion, see \cite{Tak2022}. Fig.~\ref{fig:uhdm} shows our upper limits for two annihilation channels (blue solid lines), $\tau^+\tau^-$ and $b\bar{b}$, as well as the theoretical bounds: the standard thermal-relic limit (red solid), the unitarity limit for a point-like particle respecting the partial-wave unitarity bound (purple solid), and the unitarity limits for a composite particle (purple dashed lines). Note that we assume $v_{\rm rel}$/\textit{c} = 2$\times10^{-5}$ for the relative velocity between dark matter particles in dSph galaxies~\citep{Martinez:2010xn,McGaugh:2021tyj}. Our results not only constrain part of the allowed region of a point-like dark matter cross section, but also limit the radius of UHDM in a mass range from about 100 TeV to 30 PeV. This is visible from Fig.~\ref{fig:uhdm} and depicted explicitly in Fig.~\ref{fig:uhdmR}. For example, below a dark matter mass of approximately 1 PeV, an UHDM model with the UHDM particle size of 0.6~fm or larger can be rejected at the 95\% confidence level in all annihilation channels.

Fig.~\ref{fig:comparison} shows our upper limits compared with results from the \textit{Fermi}-LAT, MAGIC, VERITAS, H.E.S.S., and HAWC collaborations. Since we use the previously published VERITAS observations, our results are similar to the published ones below 100 TeV, with the differences coming entirely from the method of extracting upper limits. Our results extend limits on the dark matter velocity-weighted annihilation cross section into a mass range which has not previously been explored. 
 
In this paper, we have presented an indirect search for an UHDM annihilation signal, using previously published VERITAS observations to access a novel dark matter parameter space. We search for final-state $\gamma$-rays from nine annihilation channels, using 216 hours of observations of four dwarf spheroidal galaxies: Segue 1, Ursa Minor, Bo\"otes, and Draco. In the absence of a detection, we have shown upper limits on the dark matter velocity-weighted annihilation cross section for dark matter particle masses from 1 TeV to 30 PeV with a joint-fit MLE analysis. This work has reported a new UHDM search with IACT observations, detailed a robust method for such searches, and should provide insight for future UHDM studies with the deep observations from the current IACTs and/or the future sensitive observatories such as the Cherenkov Telescope Array~\citep{CTA}.

\begin{acknowledgments}
{\it Acknowledgments.}
This research is supported by grants from the U.S. Department of Energy Office of Science, the U.S. National Science Foundation and the Smithsonian Institution, by NSERC in Canada, and by the Helmholtz Association in Germany. This research used resources provided by the Open Science Grid, which is supported by the National Science Foundation and the U.S. Department of Energy's Office of Science, and resources of the National Energy Research Scientific Computing Center (NERSC), a U.S. Department of Energy Office of Science User Facility operated under Contract No. DE-AC02-05CH11231. We acknowledge the excellent work of the technical support staff at the Fred Lawrence Whipple Observatory and at the collaborating institutions in the construction and operation of the instrument. D. Tak and E. Pueschel acknowledge the Young Investigators Program of the Helmholtz Association, and thank Alex Geringer-Sameth, Savvas M. Koushiappas, and Matthew Walker for providing the parameter sets for the $J$-factors. D. Tak is supported by the National Research Foundation of Korea (NRF) grant, No. 2021M3F7A1084525, funded by the Korea government (MSIT). M. Baumgart is supported by the DOE (HEP) Award DE-SC0019470.
\end{acknowledgments}

\bibliography{references}

\begin{thebibliography}{}
\expandafter\ifx\csname natexlab\endcsname\relax\def\natexlab#1{#1}\fi
\providecommand{\url}[1]{\href{#1}{#1}}

\bibitem[{Abdalla {et~al.}(2018)}]{dm_hess}
Abdalla, H., {et~al.} 2018, JCAP, 2018, 037

\bibitem[{Abdallah {et~al.}(2020)}]{dm_hess2}
Abdallah, H., {et~al.} 2020, \prd, 102, 062001

\bibitem[{Acciari {et~al.}(2022)}]{dm_magic2}
Acciari, V.~A., {et~al.} 2022, Physics of the Dark Universe, 35, 100912

\bibitem[{{Acharya} {et~al.}(2019)}]{CTA}
{Acharya}, B.~S., {et~al.} 2019, {Science with the Cherenkov Telescope Array},
  doi:10.1142/10986

\bibitem[{Ackermann {et~al.}(2015)}]{dm_fermi}
Ackermann, M., {et~al.} 2015, \prl, 115, 231301

\bibitem[{Albert {et~al.}(2018)}]{dm_hawc}
Albert, A., {et~al.} 2018, \apj, 853, 154

\bibitem[{Albert {et~al.}(2020)}]{dm_hawc2}
---. 2020, \prd, 101, 103001

\bibitem[{Aleksi{\'c} {et~al.}(2014)}]{dm_magic}
Aleksi{\'c}, J., {et~al.} 2014, JCAP, 2014, 008

\bibitem[{Archambault {et~al.}(2017)}]{dm_veritas}
Archambault, S., {et~al.} 2017, \prd, 95, 082001

\bibitem[{{Bauer} {et~al.}(2021){Bauer}, {Rodd}, \& {Webber}}]{HDM}
{Bauer}, C.~W., {Rodd}, N.~L., \& {Webber}, B.~R. 2021, Journal of High Energy
  Physics, 2021, 121

\bibitem[{{Berlin} {et~al.}(2016){Berlin}, {Hooper}, \& {Krnjaic}}]{Berlin2016}
{Berlin}, A., {Hooper}, D., \& {Krnjaic}, G. 2016, Physics Letters B, 760, 106

\bibitem[{{Bonnivard} {et~al.}(2016){Bonnivard}, {Maurin}, \&
  {Walker}}]{Bonnivard2016}
{Bonnivard}, V., {Maurin}, D., \& {Walker}, M.~G. 2016, MNRAS, 462, 223

\bibitem[{Bottaro {et~al.}(2022)Bottaro, Buttazzo, Costa, Franceschini, Panci,
  Redigolo, \& Vittorio}]{Bottaro:2021snn}
Bottaro, S., Buttazzo, D., Costa, M., {et~al.} 2022, Eur. Phys. J. C, 82, 31

\bibitem[{{Carney} {et~al.}(2022){Carney}, {Raj}, {Bai}, {Berger}, {Blanco},
  {Bramante}, {Cappiello}, {Dutra}, {Ebadi}, {Engel}, {Kolb}, {Harding},
  {Kumar}, {Krnjaic}, {Lang}, {Leane}, {Lehmann}, {Li}, {Long}, {Mohlabeng},
  {Olcina}, {Pueschel}, {Rodd}, {Rott}, {Sengupta}, {Shakya}, {Walsworth}, \&
  {Westerdale}}]{snowmass2022}
{Carney}, D., {Raj}, N., {Bai}, Y., {et~al.} 2022, arXiv e-prints,
  arXiv:2203.06508

\bibitem[{{Cirelli} {et~al.}(2011){Cirelli}, {Corcella}, {Hektor}, {H{\"u}tsi},
  {Kadastik}, {Panci}, {Raidal}, {Sala}, \& {Strumia}}]{PPPC}
{Cirelli}, M., {Corcella}, G., {Hektor}, A., {et~al.} 2011, JCAP, 2011, 051

\bibitem[{Cogan(2007)}]{VEGAS}
Cogan, P. 2007, 30th International Cosmic Ray Conference, Merida, Mexico

\bibitem[{{Contino} {et~al.}(2019){Contino}, {Mitridate}, {Podo}, \&
  {Redi}}]{Contino2019}
{Contino}, R., {Mitridate}, A., {Podo}, A., \& {Redi}, M. 2019, Journal of High
  Energy Physics, 2019, 187

\bibitem[{{Crocker} {et~al.}(2022){Crocker}, {Macias}, {Mackey}, {Krumholz},
  {Ando}, {Horiuchi}, {Baring}, {Gordon}, {Venville}, {Duffy}, {Yang},
  {Aharonian}, {Hinton}, {Song}, {Ruiter}, \& {Filipovi{\'c}}}]{Crocker2022}
{Crocker}, R.~M., {Macias}, O., {Mackey}, D., {et~al.} 2022, Nature Astronomy,
  6, 1317

\bibitem[{Fomin {et~al.}(1994)}]{Fomin1994}
Fomin, V., {et~al.} 1994, in Astroparticle Physics, Vol.~2, 137--150

\bibitem[{Geller {et~al.}(2018)Geller, Iwamoto, Lee, Shadmi, \&
  Telem}]{Geller2018}
Geller, M., Iwamoto, S., Lee, G., Shadmi, Y., \& Telem, O. 2018, JHEP, 06, 135

\bibitem[{{Geringer-Sameth} {et~al.}(2015){Geringer-Sameth}, {Koushiappas}, \&
  {Walker}}]{GS2015}
{Geringer-Sameth}, A., {Koushiappas}, S.~M., \& {Walker}, M. 2015, \apj, 801,
  74

\bibitem[{{Griest} \& {Kamionkowski}(1990)}]{Griest1990}
{Griest}, K., \& {Kamionkowski}, M. 1990, \prl, 64, 615

\bibitem[{{Harigaya} {et~al.}(2016){Harigaya}, {Ibe}, {Kaneta}, {Nakano}, \&
  {Suzuki}}]{Harigaya2016}
{Harigaya}, K., {Ibe}, M., {Kaneta}, K., {Nakano}, W., \& {Suzuki}, M. 2016,
  Journal of High Energy Physics, 2016, 151

\bibitem[{{Hernquist}(1990)}]{Hernquist1990}
{Hernquist}, L. 1990, \apj, 356, 359

\bibitem[{Holder {et~al.}(2008)}]{VERITASInstrument}
Holder, J., {et~al.} 2008, AIP Conference Proceedings, 1085, 657.
\newblock \url{https://aip.scitation.org/doi/abs/10.1063/1.3076760}

\bibitem[{{Kieda} {et~al.}(2013)}]{Kieda2013}
{Kieda}, D., {et~al.} 2013, in International Cosmic Ray Conference, Vol.~33,
  33rd International Cosmic Ray Conference (ICRC2013), 1124

\bibitem[{{Li} \& {Ma}(1983)}]{Li1983}
{Li}, T.~P., \& {Ma}, Y.~Q. 1983, \apj, 272, 317

\bibitem[{Martinez {et~al.}(2011)Martinez, Minor, Bullock, Kaplinghat, Simon,
  \& Geha}]{Martinez:2010xn}
Martinez, G.~D., Minor, Q.~E., Bullock, J., {et~al.} 2011, Astrophys. J., 738,
  55

\bibitem[{McGaugh {et~al.}(2021)McGaugh, Lelli, Schombert, Li, Visgaitis,
  Parker, \& Pawlowski}]{McGaugh:2021tyj}
McGaugh, S.~S., Lelli, F., Schombert, J.~M., {et~al.} 2021, Astron. J., 162,
  202

\bibitem[{{Park} {et~al.}(2015)}]{Park2015}
{Park}, N., {et~al.} 2015, in International Cosmic Ray Conference, Vol.~34,
  34th International Cosmic Ray Conference (ICRC2015), 771

\bibitem[{Tak {et~al.}(2022)Tak, Baumgart, Rodd, \& Pueschel}]{Tak2022}
Tak, D., Baumgart, M., Rodd, N.~L., \& Pueschel, E. 2022, ApJL, 938, L4.
\newblock \url{https://dx.doi.org/10.3847/2041-8213/ac9387}

\bibitem[{{Zhao}(1996)}]{Zhao1996}
{Zhao}, H. 1996, MNRAS, 278, 488

\end{thebibliography}

\appendix

\section{Significance of the dark matter annihilation signal}\label{sec:sig}
\begin{figure*}[b!]
    \centering
    \subfigure{\includegraphics[width=0.3\linewidth]{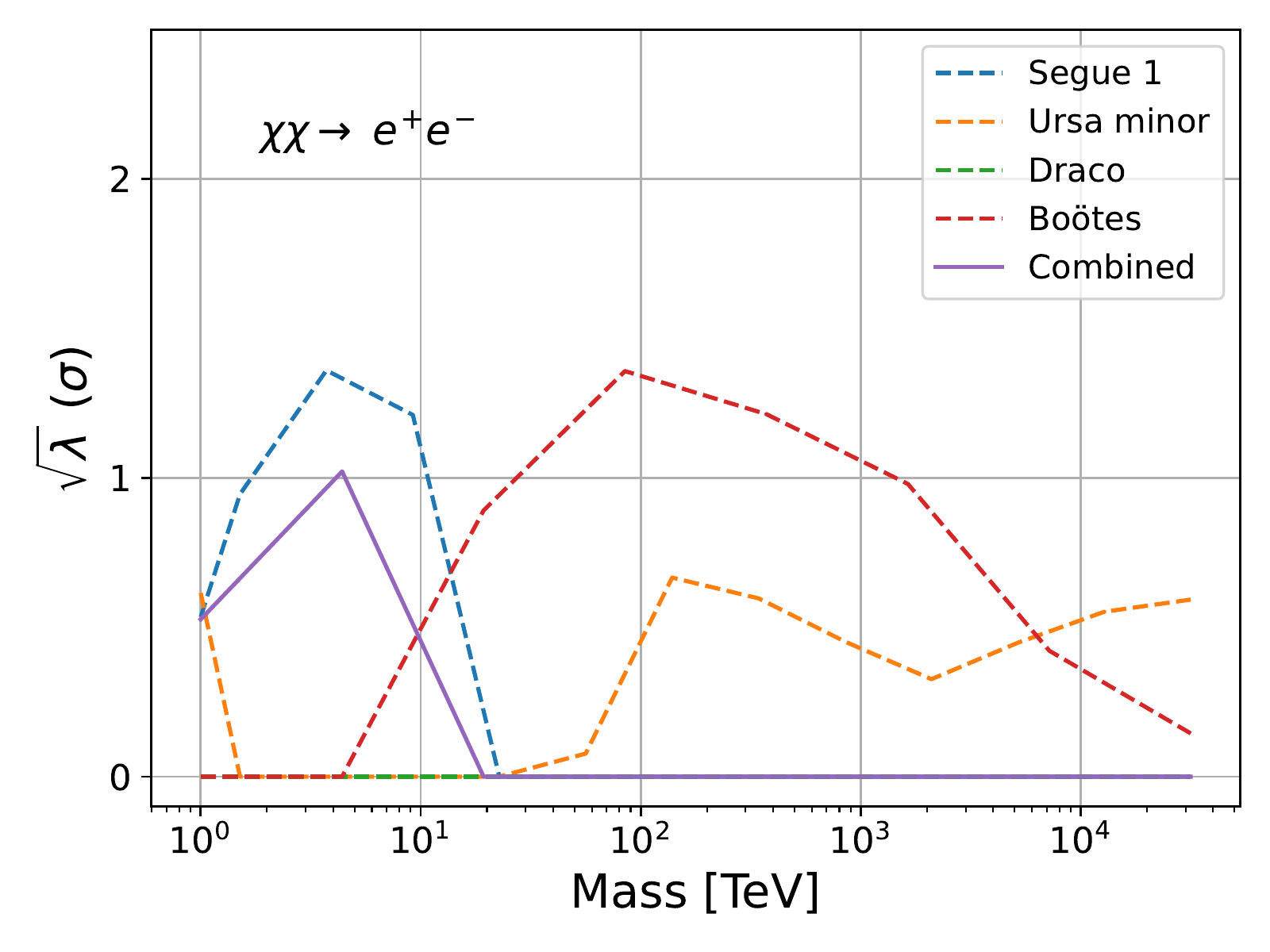}}
    \subfigure{\includegraphics[width=0.3\linewidth]{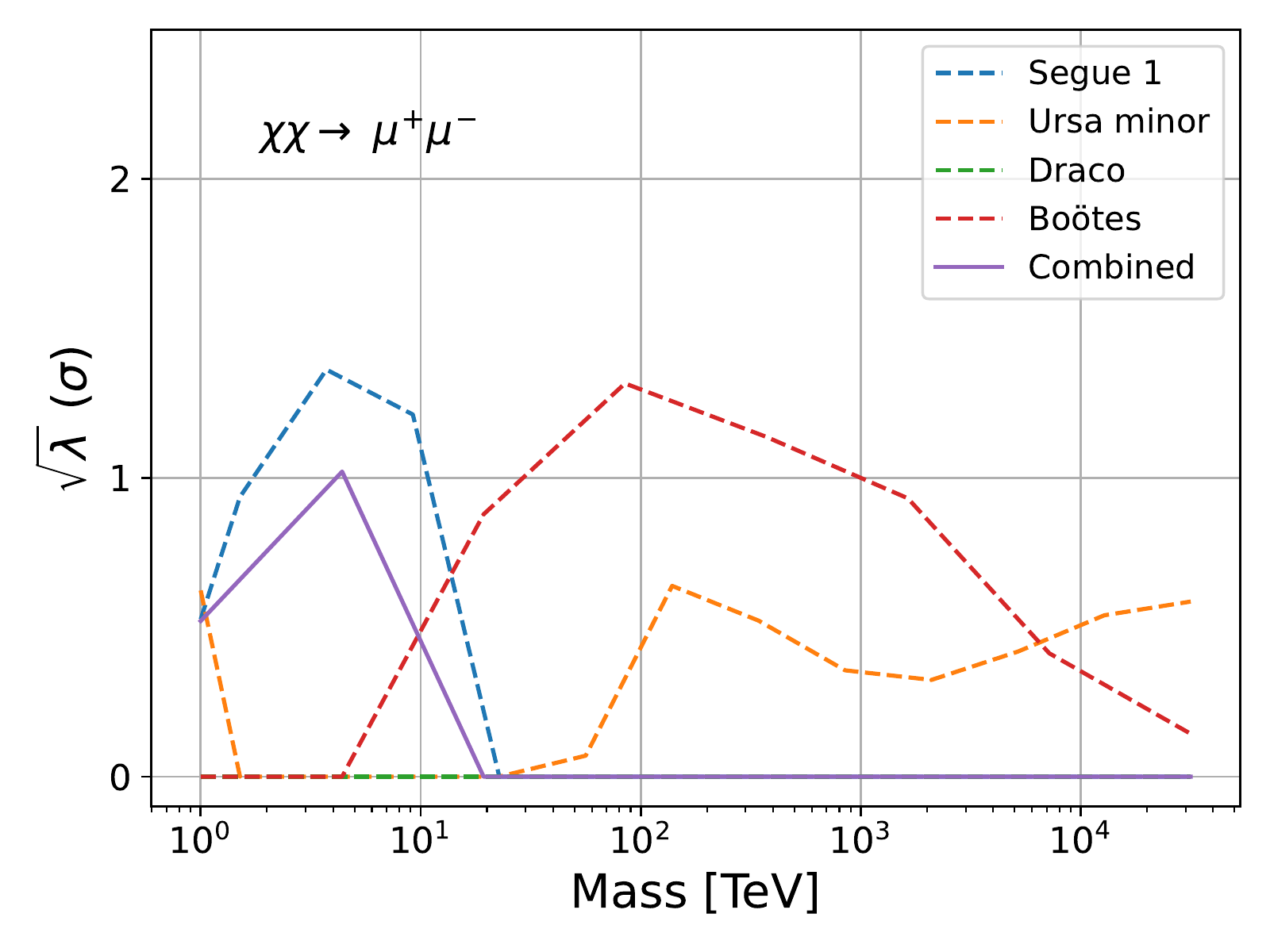}}
    \subfigure{\includegraphics[width=0.3\linewidth]{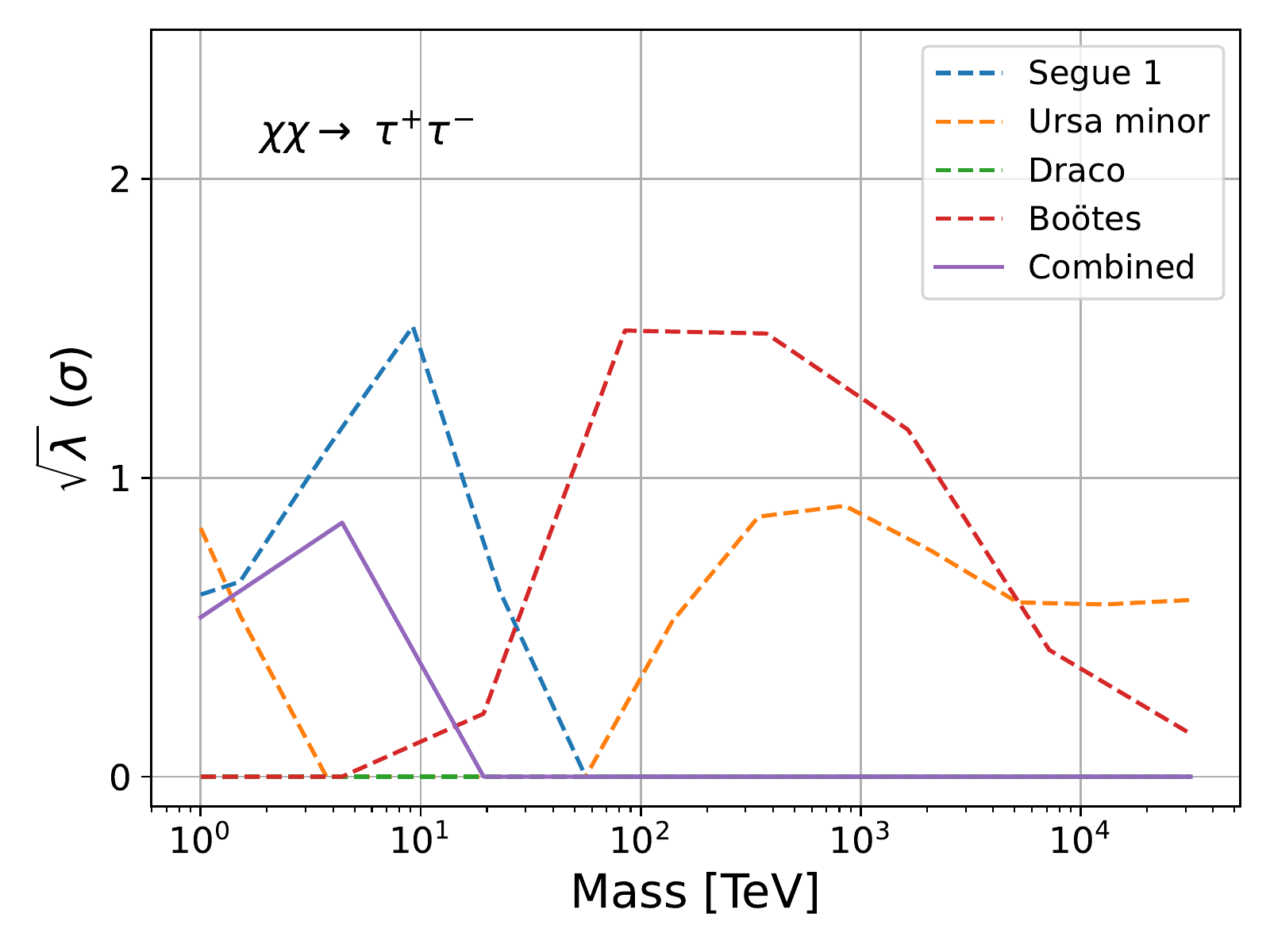}}\\
    \subfigure{\includegraphics[width=0.3\linewidth]{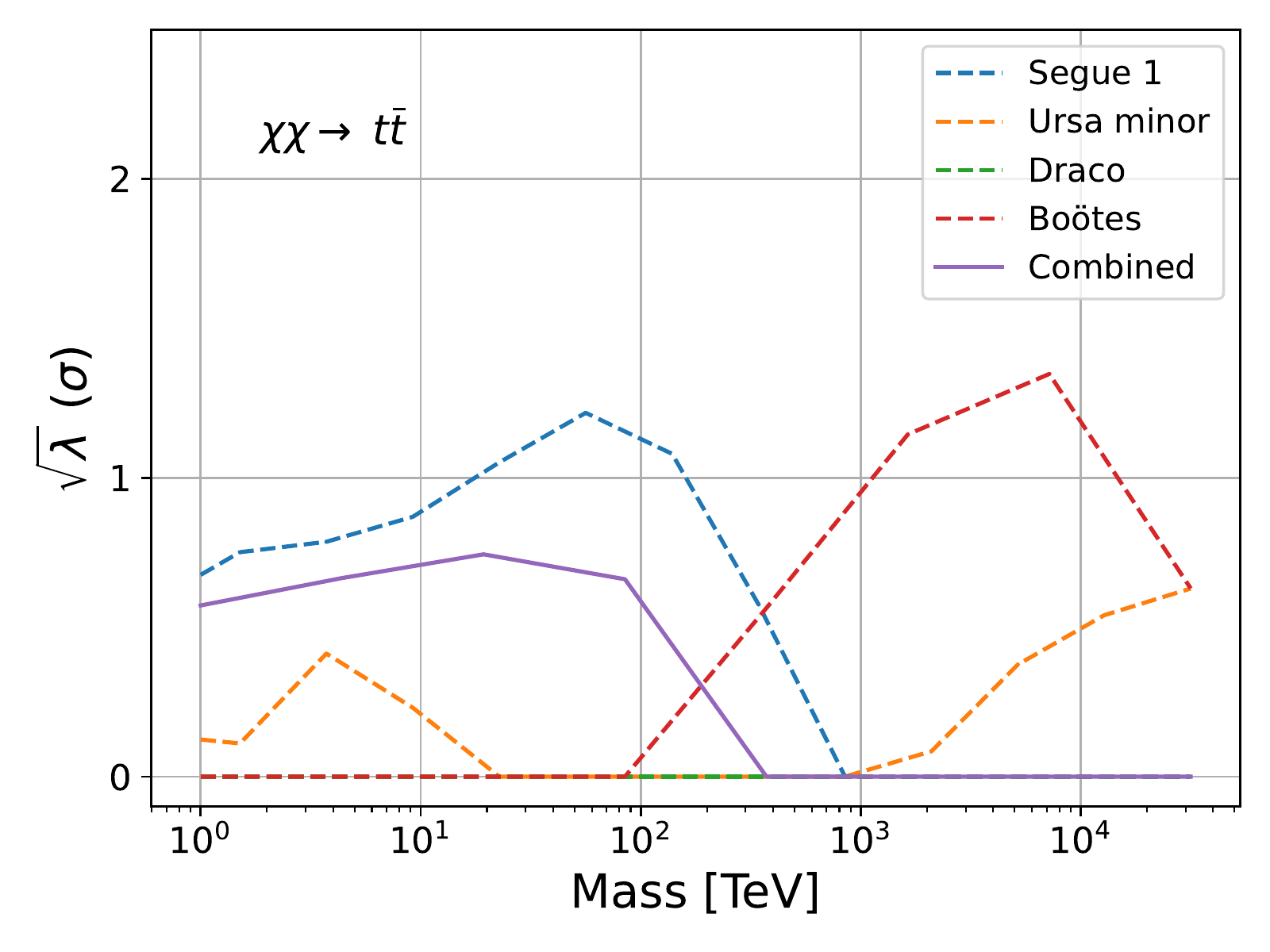}}
    \subfigure{\includegraphics[width=0.3\linewidth]{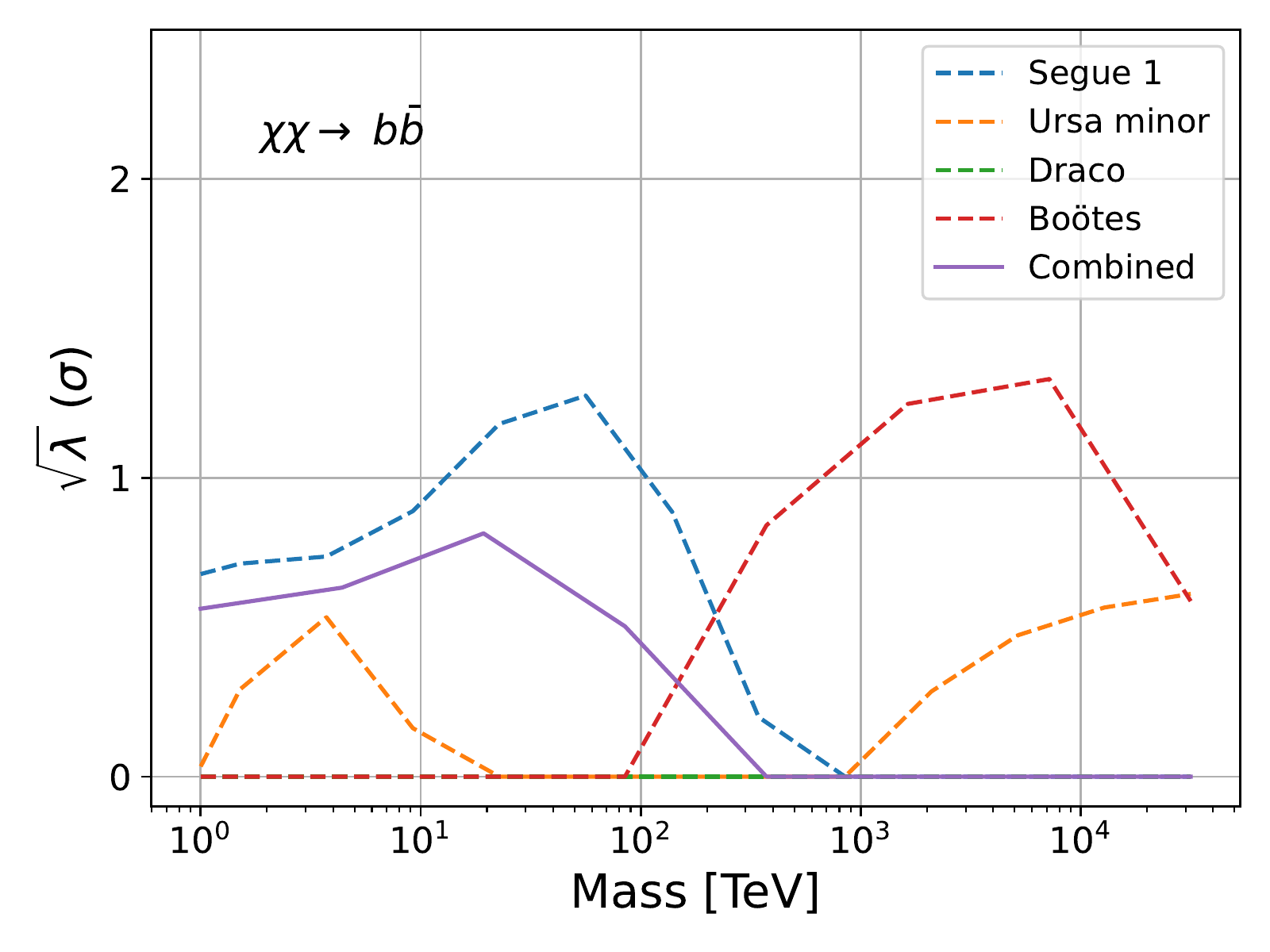}}
    \subfigure{\includegraphics[width=0.3\linewidth]{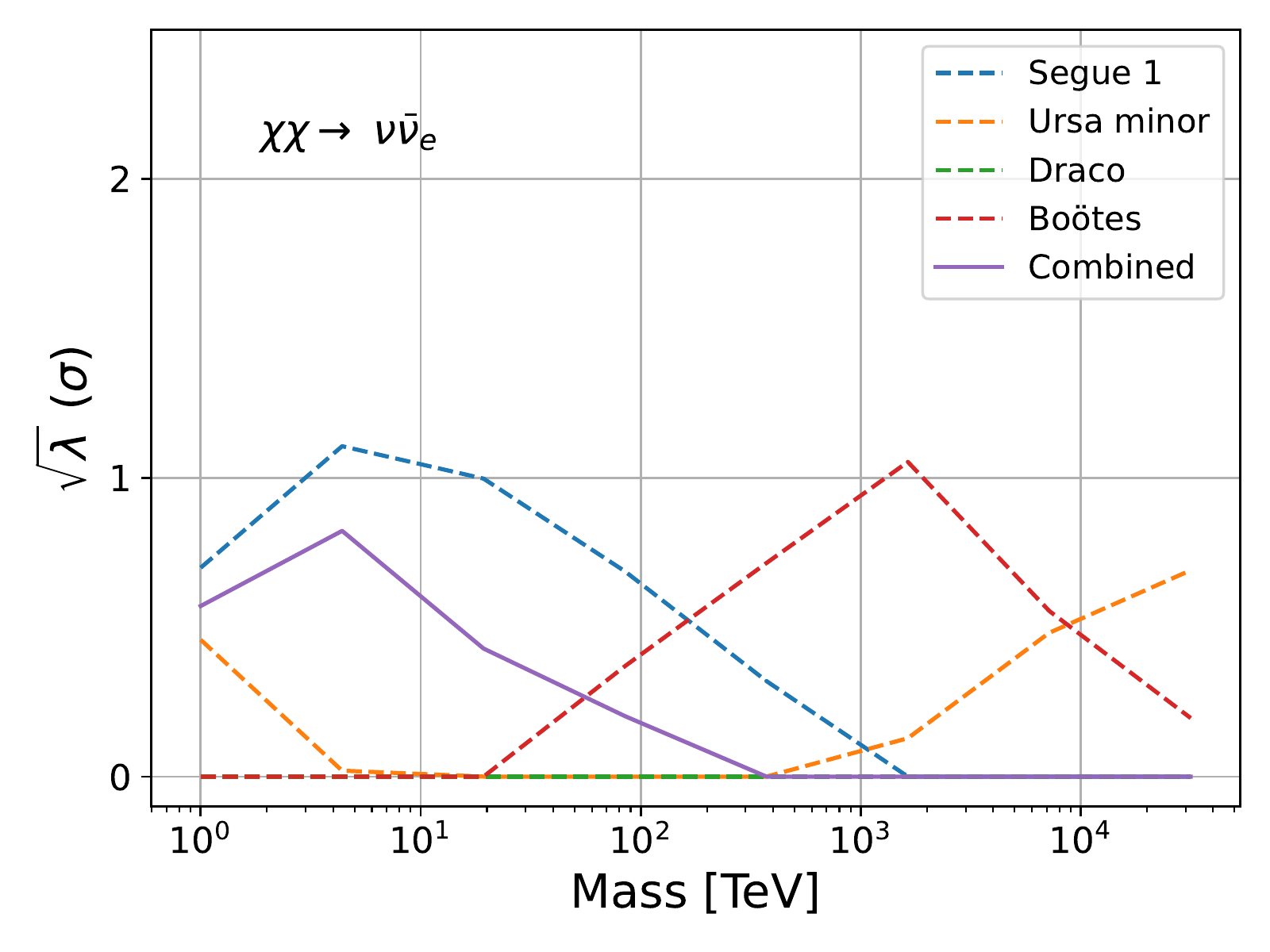}}\\
    \subfigure{\includegraphics[width=0.3\linewidth]{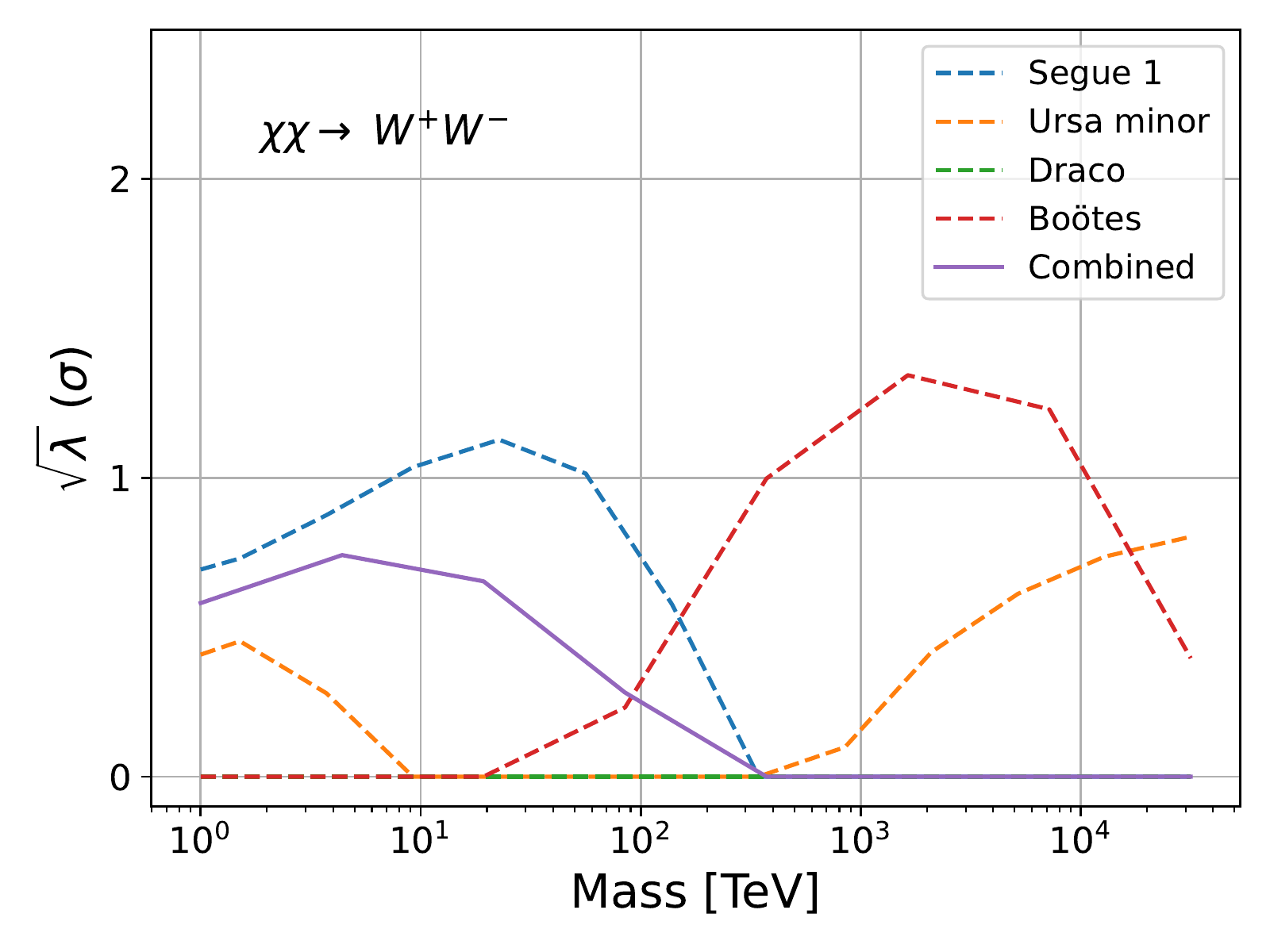}}
    \subfigure{\includegraphics[width=0.3\linewidth]{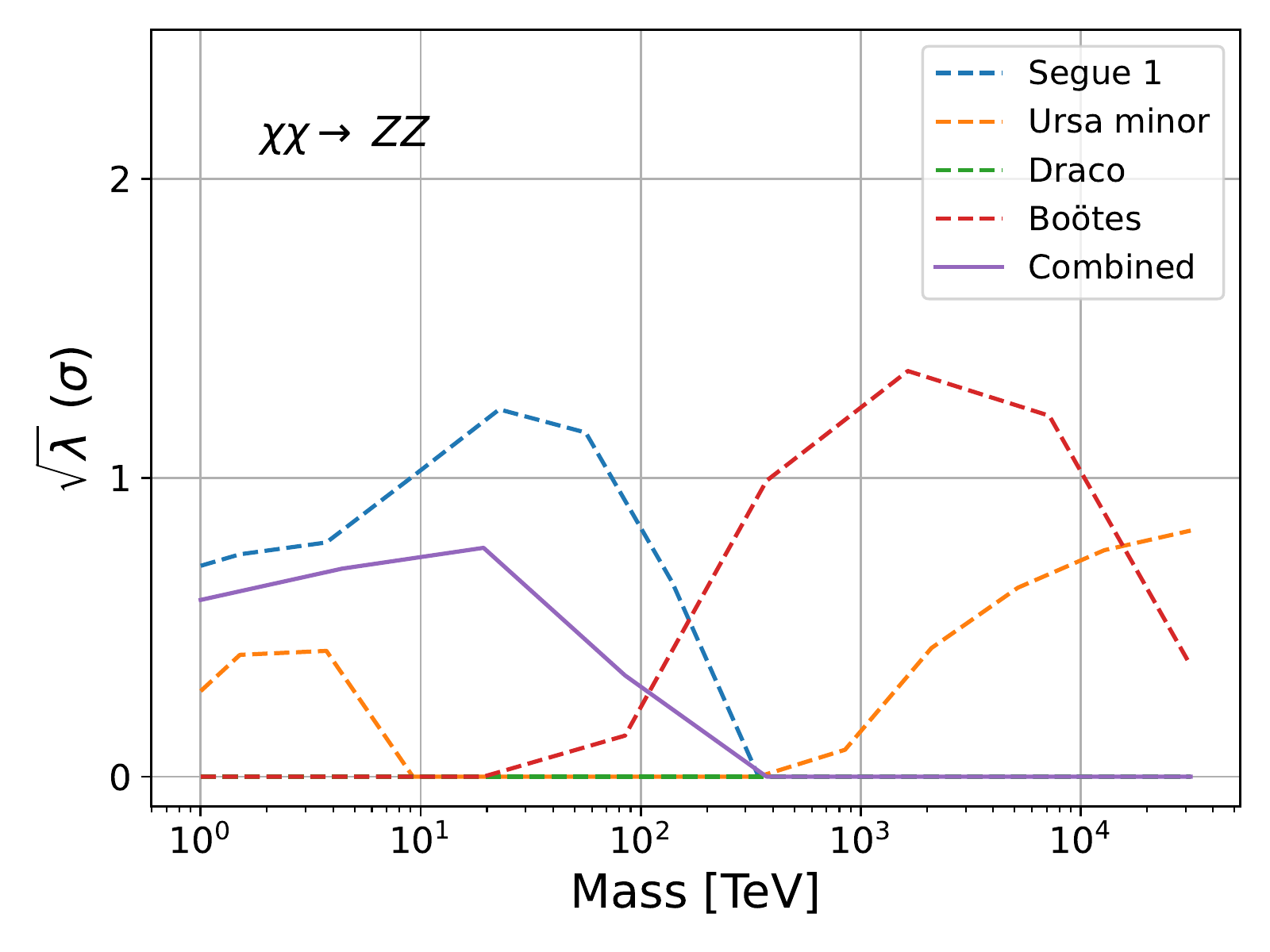}}
    \subfigure{\includegraphics[width=0.3\linewidth]{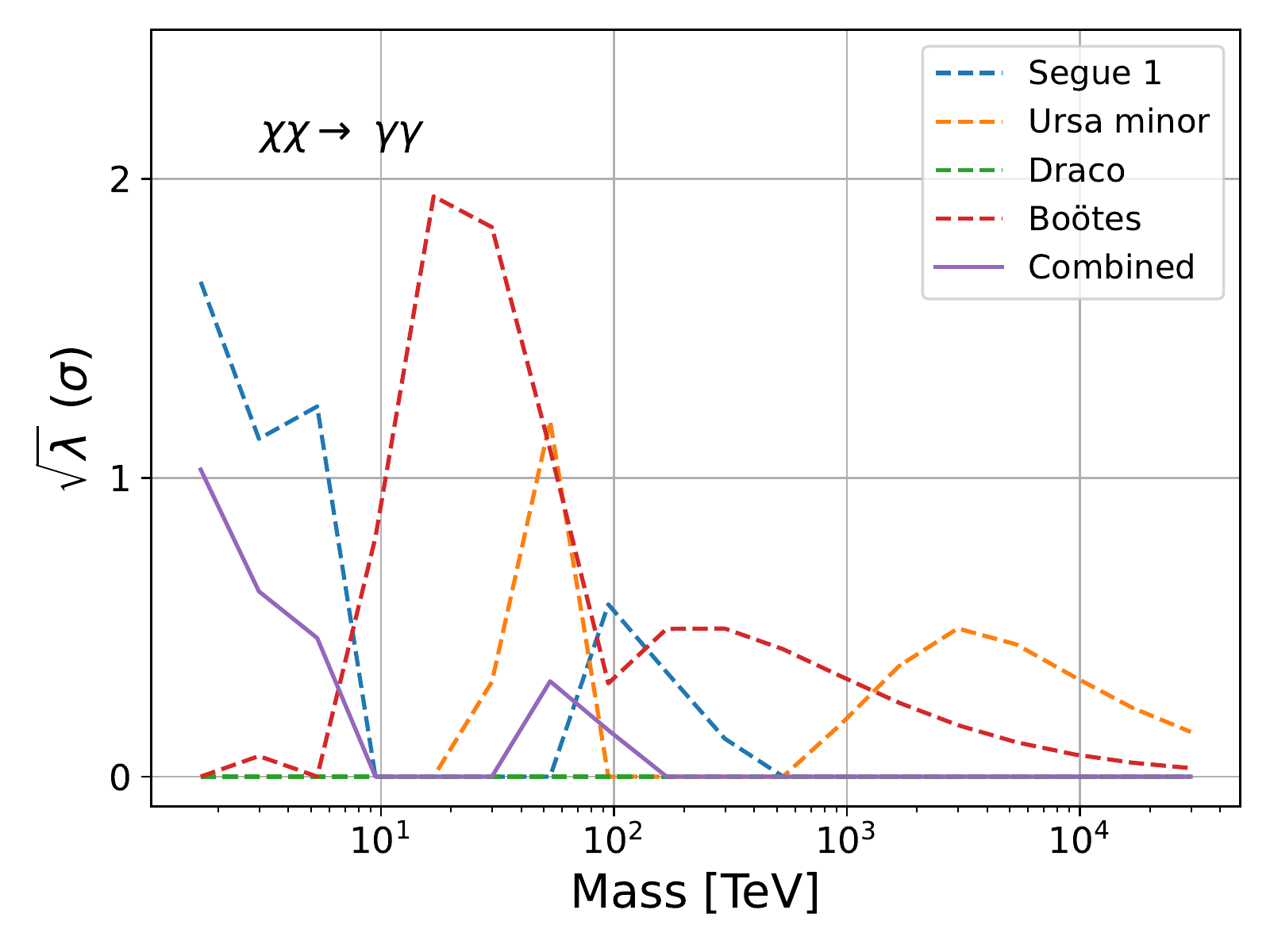}}
    \caption{VERITAS-measured significances of the dark matter annihilation signal in nine annihilation channels for the individual dSphs and for their combination. The dashed lines show the signal significance as a function of dark matter particle mass. The solid curve shows the combined significance.}
    \label{fig:ap_sig}
\end{figure*}

Fig.~\ref{fig:ap_sig} shows the signal significance (given by $\sqrt{\lambda}$) as a function of dark matter particle mass in the nine annihilation channels. The significance curves for the individual dSphs are shown, as well as the combined results. For no dark matter particle mass, dSph, or annihilation channel does the signal significance reach 2$\sigma$.

We note that the significance in Fig.~\ref{fig:ap_sig} is calculated from the likelihood analysis with observed ON and OFF regions. This result shows the non-detection of a DM signal. Fig.~\ref{fig:exp_uls}, in contrast, compares observed upper limits with expected upper limit bands assuming a simulated ON region made up of randomly sampled observed OFF-region events. The observed agreement between the observed limits and expected limit band implies that observed ON region is consistent with the Poisson fluctuation of observed OFF regions. These two approaches lead to the same conclusion that we do not observe any excess (a possible dark matter signal) from our observations.

\section{Upper limit curves from the four dwarf spheroidal galaxies}\label{sec:uls}
\begin{figure*}[t!]
    \centering
    \subfigure{\includegraphics[width=0.3\linewidth]{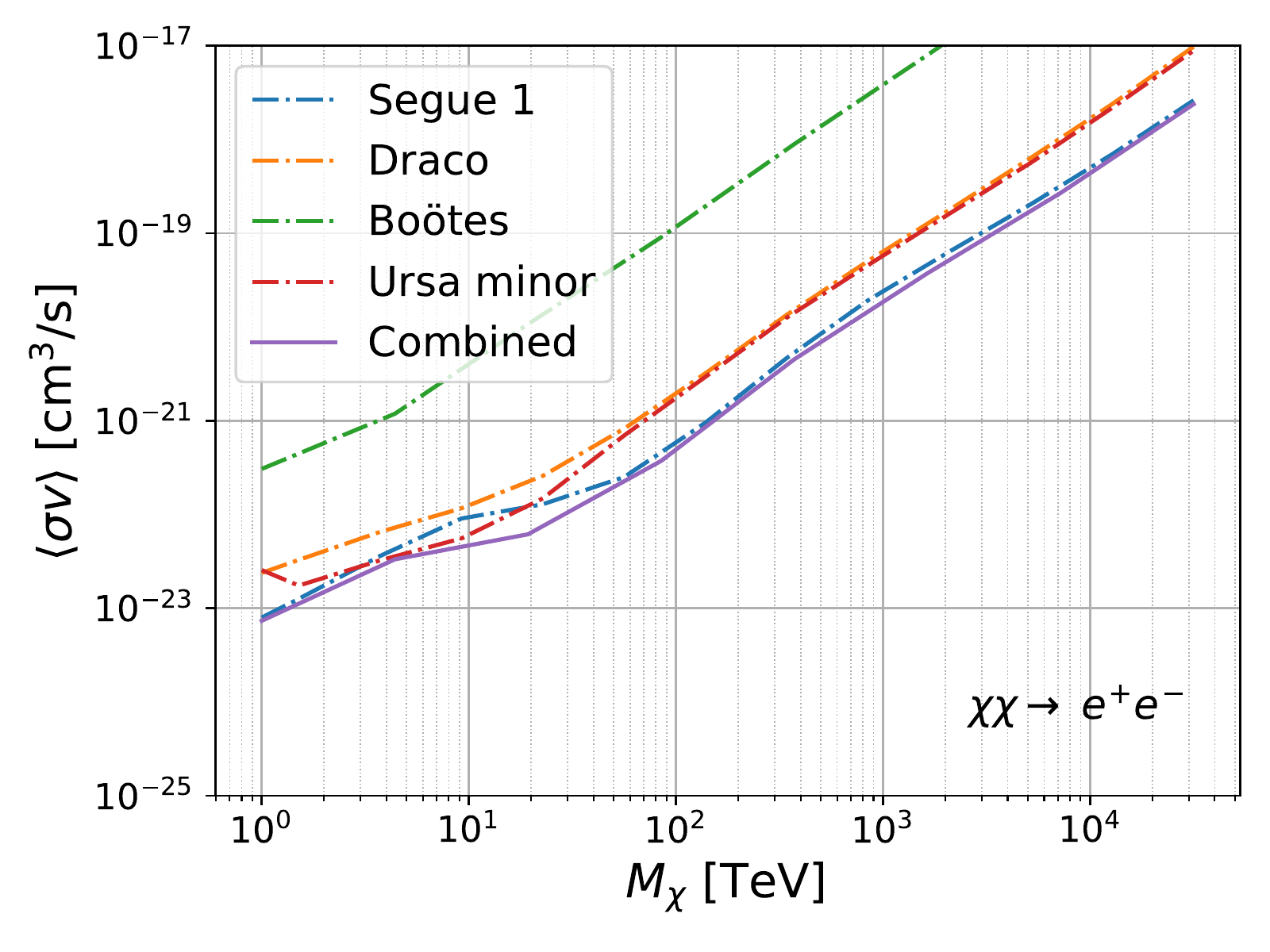}}
    \subfigure{\includegraphics[width=0.3\linewidth]{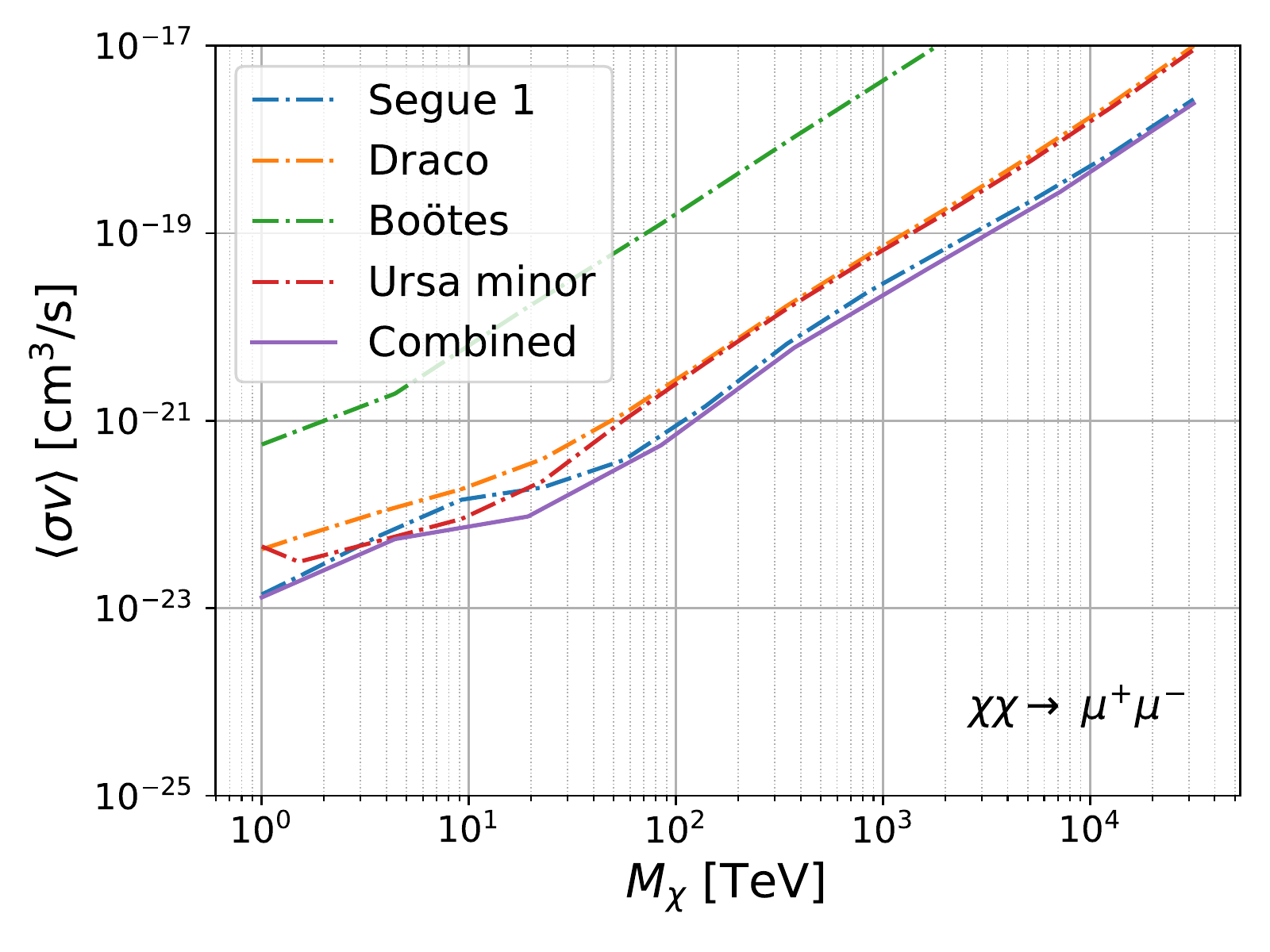}}
    \subfigure{\includegraphics[width=0.3\linewidth]{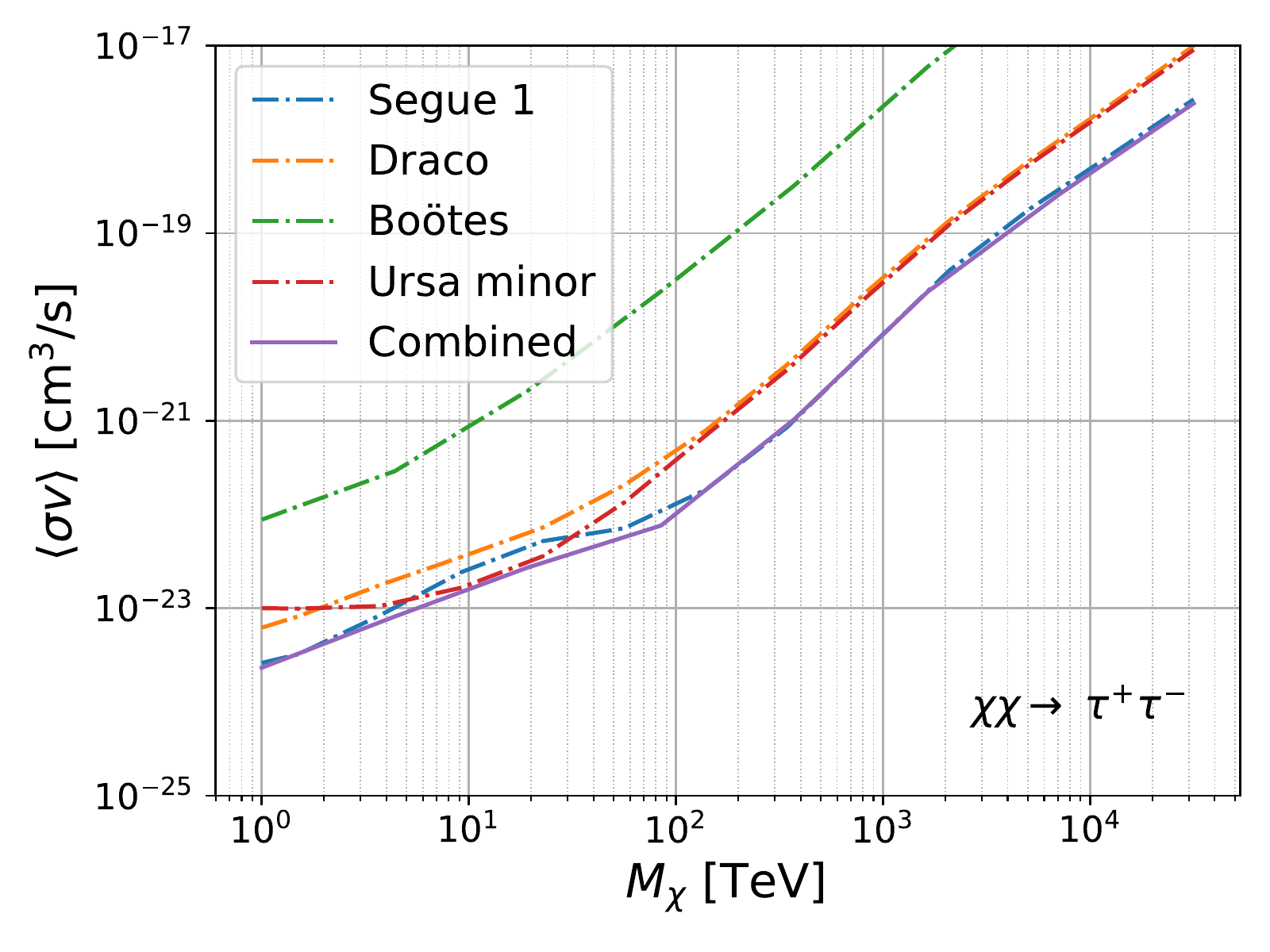}}\\
    \subfigure{\includegraphics[width=0.3\linewidth]{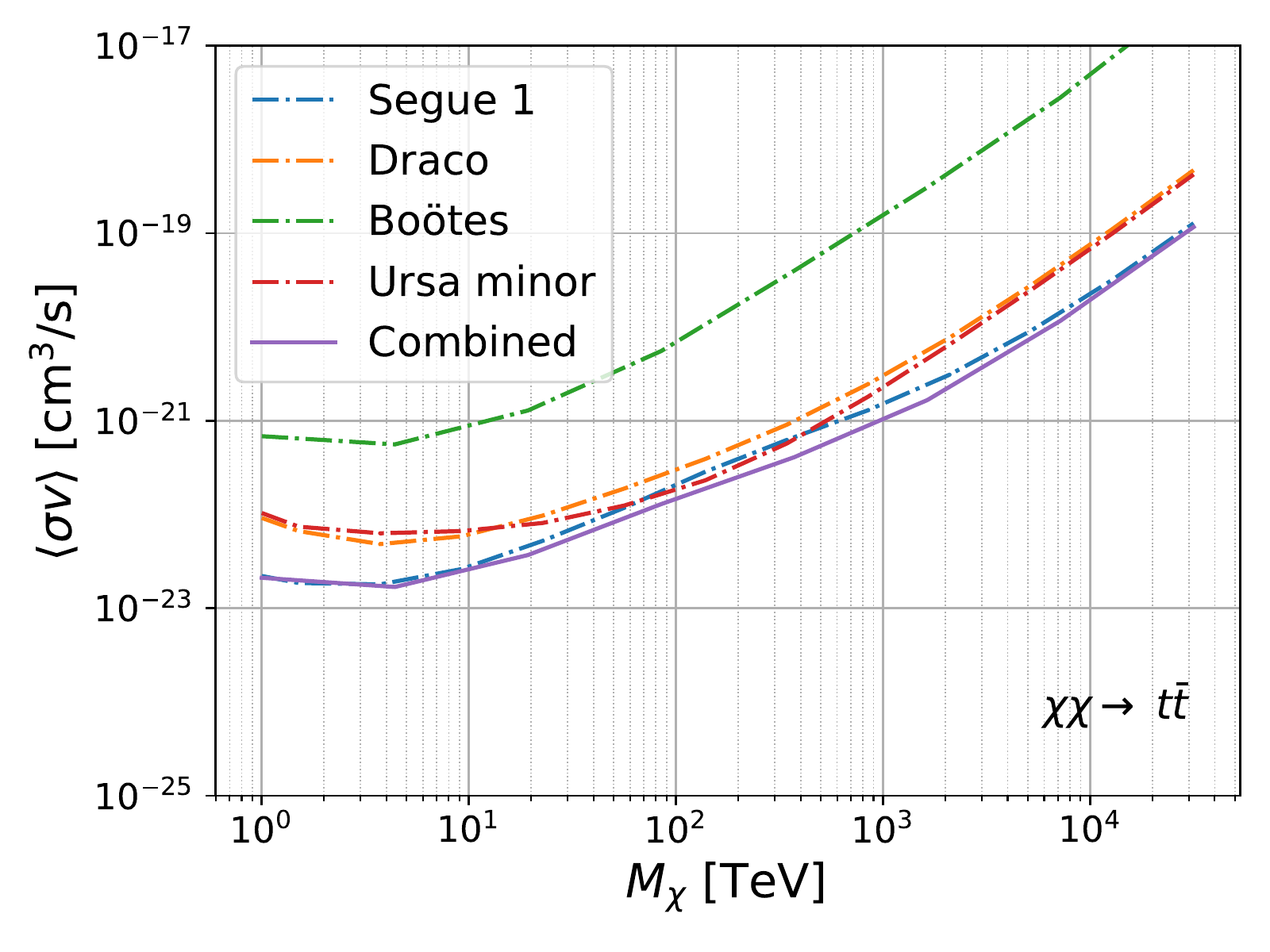}}
    \subfigure{\includegraphics[width=0.3\linewidth]{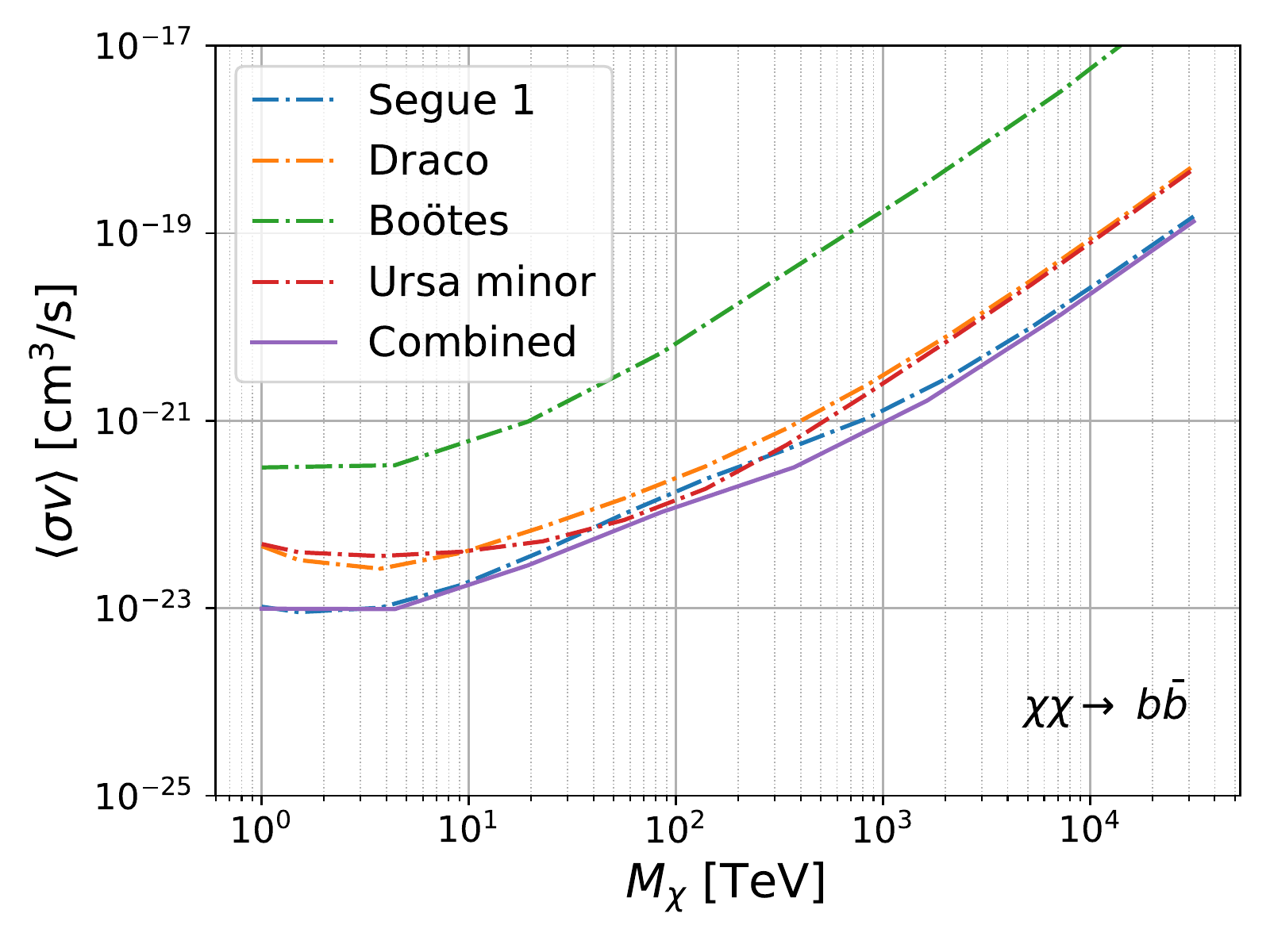}}
    \subfigure{\includegraphics[width=0.3\linewidth]{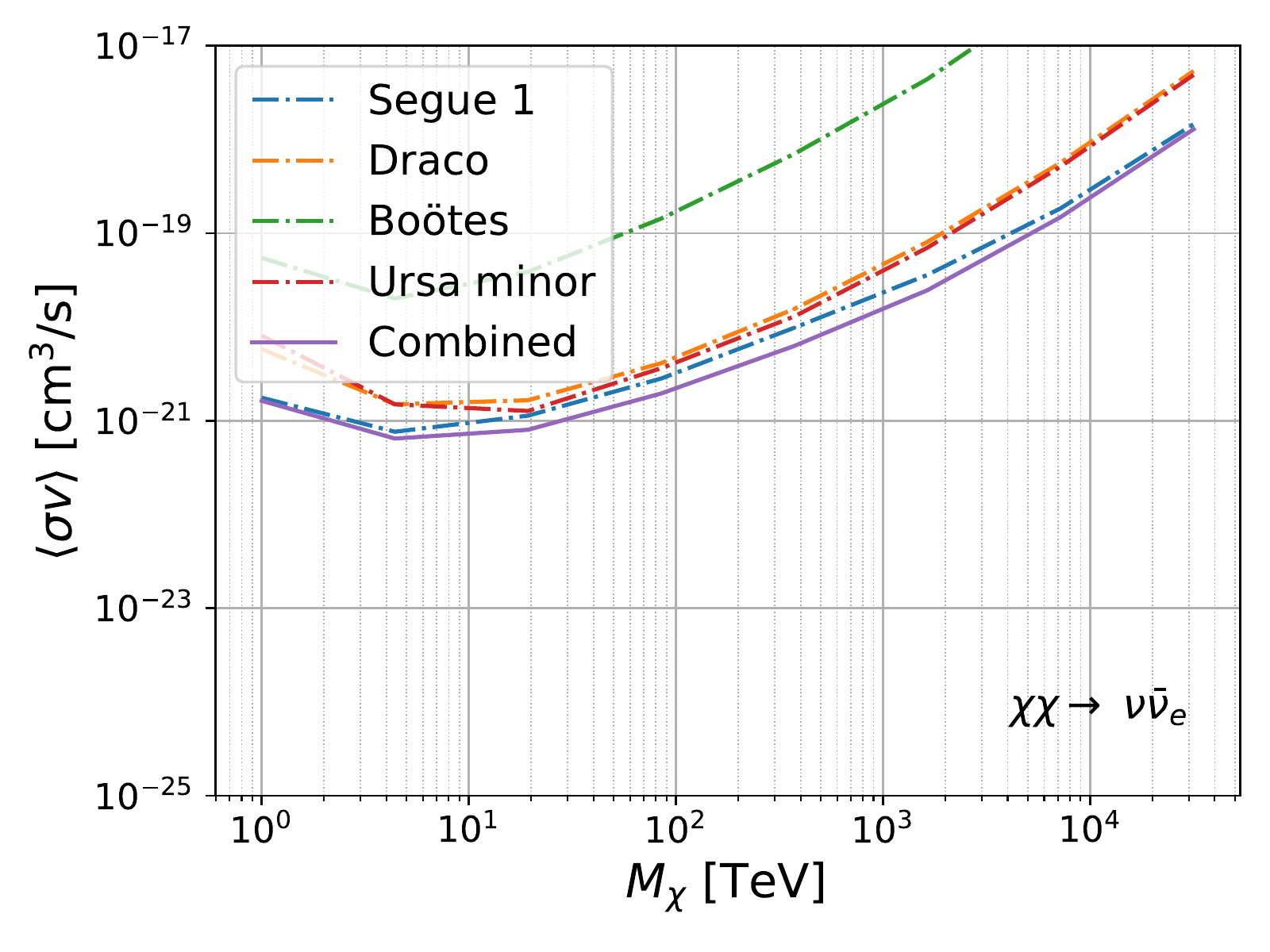}}\\
    \subfigure{\includegraphics[width=0.3\linewidth]{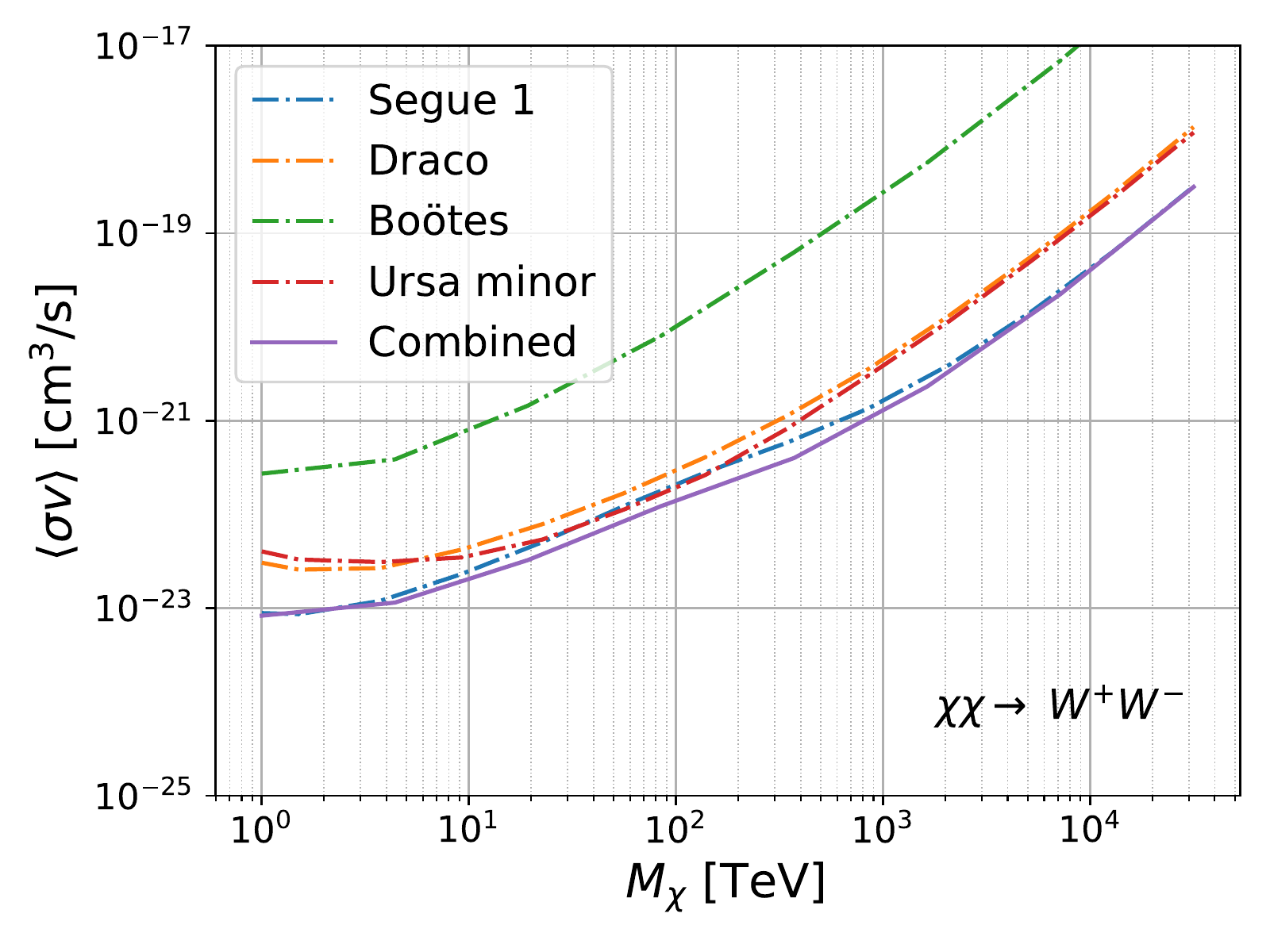}}
    \subfigure{\includegraphics[width=0.3\linewidth]{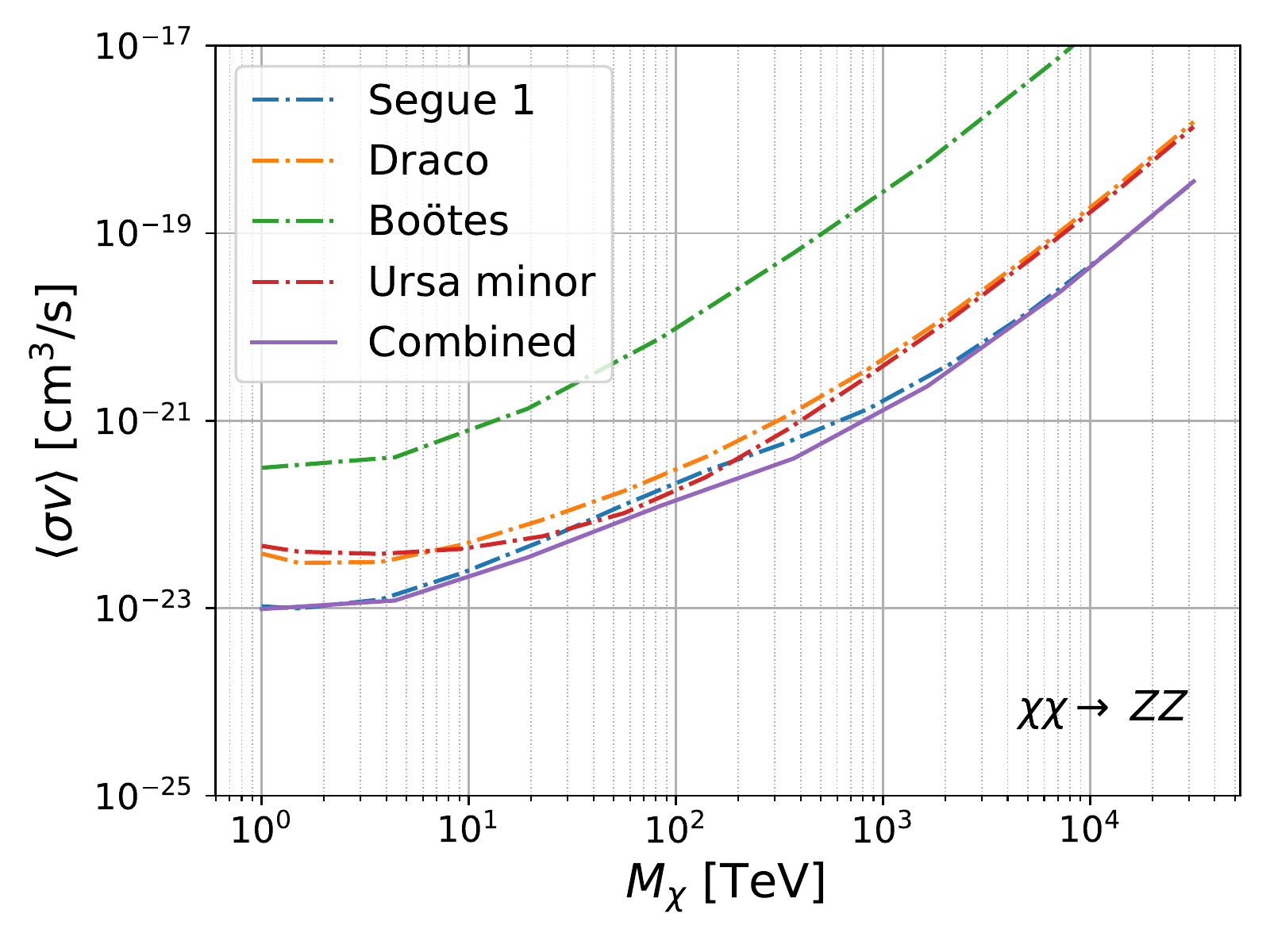}}
    \subfigure{\includegraphics[width=0.3\linewidth]{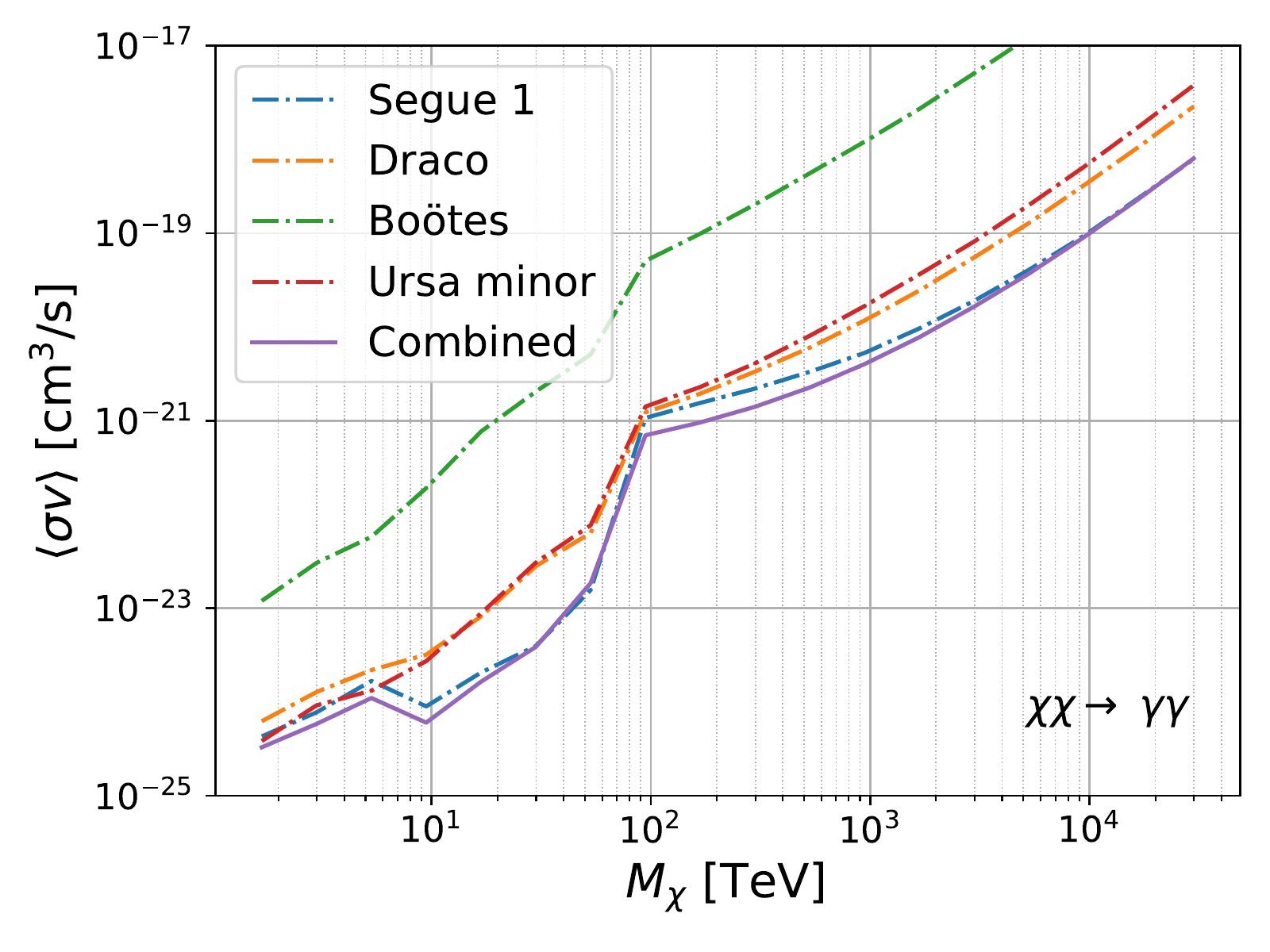}}
    \caption{VERITAS upper limits derived from observations of the four dSphs, considering nine annihilation channels. The dot-dashed lines indicate the limits from the individual dSphs, while the solid lines indicate the combined limits.}
    \label{fig:ap_uls}
\end{figure*}

Fig.~\ref{fig:ap_uls} shows the upper limits on the UHDM velocity-weighted annihilation cross section as a function of particle mass for each dSph considered, as well as the combined limit. As in the main text, nine annihilation channels are considered. As expected based on the $J$-factors listed in Table~\ref{tab:nfw}, Segue 1 generally provides the most constraining limits, followed by Ursa Minor and Draco, with the weakest limits coming from Bo\"otes.

\end{document}